\documentclass[12pt]{article}
\usepackage[left=2cm, right=2cm, bottom=2cm ,top = 2cm, footskip=1cm]{geometry}
\usepackage[font=small,labelfont=bf]{caption}
\usepackage[hidelinks]{hyperref}

\usepackage{setspace}
\usepackage{changepage}

\usepackage{lineno}
\usepackage{chngcntr}

\usepackage[numbers]{natbib}
\setcitestyle{open={[},close={]},italic}

\usepackage{graphicx}
\usepackage{amsmath}

\usepackage[section]{placeins}

\title{Neuro-evolutionary evidence for a universal fractal primate brain shape}

\author{Yujiang Wang$^{1,2,3*}$, Karoline Leiberg$^{1}$, Nathan Kindred $^{2}$, Christopher R. Madan$^{4}$,\\ Colline Poirier$^{2}$, Christopher I. Petkov$^{2,5}$, \\Peter N. Taylor$^{1,2,3}$, Bruno Mota$^{6}$}

\begin{document}

\maketitle

% author affiliations
\begin{enumerate}
\item{CNNP Lab (www.cnnp-lab.com), Interdisciplinary Computing and Complex BioSystems Group, School of Computing, Newcastle University, Newcastle upon Tyne, United Kingdom}
\item{Faculty of Medical Sciences, Newcastle University, Newcastle upon Tyne, United Kingdom}
\item{UCL Institute of Neurology, Queen Square, London, United Kingdom}
\item{School of Psychology, University of Nottingham, Nottingham, United Kingdom}
\item{Department of Neurosurgery, University of Iowa, USA}
\item{metaBIO Lab, Instituto de Física, Universidade Federal do Rio de Janeiro (UFRJ), Rio de Janeiro, Brazil}
\end{enumerate}

\begin{center}
* Yujiang.Wang@newcastle.ac.uk    
\end{center}


\textbf{Classification:} PHYSICAL SCIENCES, Biophysics and Computational Biology, Evolution, Neuroscience

\newpage
\section*{Abstract}
The cerebral cortex displays a bewildering diversity of shapes and sizes across and within species. Despite this diversity, we present a universal multi-scale description of primate cortices. We show that all cortical shapes can be described as a set of nested folds of different sizes. As neighbouring folds are gradually merged, the cortices of 11 primate species follow a common scale-free morphometric trajectory, that also overlaps with over 70 other mammalian species. Our results indicate that all cerebral cortices are approximations of the \textit{same} archetypal fractal shape with a fractal dimension of $d_f=2.5$. Importantly, this new understanding enables a more precise quantification of brain morphology as a function of scale. To demonstrate the importance of this new understanding, we show a scale-dependent effect of ageing on brain morphology. We observe a more than four-fold increase in effect size (from 2 standard deviations to 8 standard deviations) at a spatial scale of approximately 2~mm compared to standard morphological analyses. Our new understanding may therefore generate superior biomarkers for a range of conditions in the future.

\newpage
\section{Introduction}

The morphological complexity of the mammalian cerebral cortex has fascinated scientists for generations, with cortices across and within species exhibiting a large diversity of shapes and sizes. Such diversity is not arbitrary, however. The mammalian brain folds into stereotypical, hierarchically-organized structures such as lobes and major gyri. In fact, qualitative and quantitative regularities in cortical scaling have often been suggested and observed  \citep{Zhang2000,Francis2009,Karbowski2011,mota_herculano-houzel_2014}. More specifically, through modelling the mechanism of cortical folding from a statistical physics approach, we have previously derived a theoretical scaling law relating pial surface area $A_t$, exposed surface area $A_e$\footnote{The exposed surface area can be thought of as the surface area of a piece of cling film wrapped around the brain. Mathematically, for the remaining paper it is the convex hull of the brain surface.}, and average cortical thickness $T$:
\begin{equation}
A_t  T^{\frac{1}{2}} = k A_e^{\frac{5}{4}}.
\label{eqn1_scalinglaw}
\end{equation}

This scaling law, relating powers of cortical thickness and surface area metrics, was shown to be valid across mammalian species  \citep{science2015} and within the human species~ \citep{pnas2016}, as well as to the structures and substructures of individual brains~ \citep{commbiol2019,leiberg2021}. Notably, across all these cases the dimensionless offset $k$ is shown to be near invariant. However, this universality, presumed to be arising from universal physical principles and evolutionarily conserved biomechanics, says little about what that cortical shape actually is, beyond a constraint binding three morphometric parameters. In this paper, we take Eqn.~(\ref{eqn1_scalinglaw}) as an empirical starting point to create a new and hierarchical way of expressing cortical shape. Specifically, we introduce a coarse-graining procedure that renders the cortex at different spatial scales, or resolutions. We show that \textit{coarse-grained primate cortices at each spatial scale can be understood as approximations of the same universal self-similar archetypal form}, of which the observed scaling law (Eqn.~\ref{eqn1_scalinglaw}) can then be shown to be a direct consequence.

Besides revealing a symmetry in nature hidden under much apparent complexity, our results indicate a conservation of morphological relationships across evolution. We will show that these results further provide us with a new and powerful tool to express and analyse cortical morphology. As an example, we will calculate the effects of human ageing across spatial scales and show that the effects are highly scale-dependent. 

\subsection{Mathematical background}
The universal scaling law (Eqn.~\ref{eqn1_scalinglaw}) can be rewritten in a suggestive way
\begin{equation}
\frac{A_t}{A_0} = \left( \frac{A_e}{A_0}\right) ^{1.25},
\label{eqn2_A0}
\end{equation}
where the $A_0 = \frac{T^2}{k^4}$ is a fundamental area element that defines the threshold between gyrencephaly (folded cortex) and lissencephaly (smooth cortex) when $A_t=A_e=A_0$. For a constant $k$ the value of $A_0$ is a multiple of $T^2$, indicating that cortical thickness determines the size of the smallest possible gyri and sulci.

This re-writing highlights a new perspective, or interpretation of the scaling law: it now suggests a relationship between intrinsic and extrinsic measures of cortical size\footnote{Given the folded laminar structure of the cortex, areas are the more natural way of measuring its ``size''}: $A_t$ and $A_e$ respectively, measured in units of $A_0$. This is reminiscent of fractal scaling \citep{Mandelbrot1983Book}, where a complex shape reveals ever smaller levels of self-similar detail as it is probed in ever smaller scales (or equivalently, higher resolutions), represented here by $A_0$. The scaling, or power exponent between the measured intrinsic and extrinsic sizes is the so-called fractal dimension. 

Although actual fractals are mathematical abstractions, they can often be defined as the limit of iterative processes. Many structures in nature, and in particular biology  \citep{Elston2005,Codling2008,IONESCU2009,Losa2011,Klonowski2016,DiIevabook2016,Reznikov2018}, are good approximations of a fractal. Equation (\ref{eqn2_A0}) is suggestive, but not proof, that cortices are among these forms, with a fractal dimension of $1.25\times2 = 2.5$ (the factor 2 being the topological dimension of areas). Indeed, fractal scaling for various aspects of cortical morphology has often been postulated  \citep{Free1996, KISELEV2003}, with a number of recent papers making use of MRI data  \citep{Marzi2021,Jao2021,Meregalli2022}. Most recently published estimates of fractal dimension for the whole cortex are indeed close to $2.5$  \citep{King2010,madan_cortical_2016, Marzi2020}.

Here, for the first time, we propose to directly construct morphologically plausible realisations of cortices at any specified spatial scale, or resolution. This is achieved through a coarse-graining method that removes morphological details smaller than a specified scale while preserving surface integrity. For example, at a set scale of 3~mm, sulcal walls that are less than 3~mm apart would be removed, and the neighbouring gyri would be fused. This method is a new systematic way of obtaining shape properties from the cortex in terms of a sequence of morphometric measurements as spatial scale varies. By examining how areas scale across coarse-grained versions of actual primate cortices, we will be able to directly verify cortical self-similarity.

\section{Method\label{methods}}

\subsection{Coarse-graining method}

As a starting point for a coarse-graining method, we suggest to turn to a well-established method that measures fractal dimension of objects: the so-called box-counting algorithm \cite{Kochunov2007,Madan2019}. Briefly, this algorithm fills the object of interest (the cortex in our case) with boxes, or voxels of increasingly larger sizes and counts the number of boxes in the object as a function of box size. As the box size increases, the number of boxes decreases; and in a log-log plot, the slope of this relationship indicates the fractal dimension of the object. In our case, this method would not only provide us with the fractal dimension of the cortex, but, with increasing box size, the filled cortex would also contain less and less detail of the folded cortex. Intuitively, with increasing box size, the smaller details below the resolution of a single box would disappear first, and increasingly larger details will follow -- precisely what we require from a coarse-graining method. We therefore propose to expand the traditional box-counting method beyond its use to measure fractal dimension, but to analyse the reconstructed cortices as different realisations of the original cortex at the specified spatial scale.

\begin{figure}[h!]
\hspace{-0.7cm}
\includegraphics[scale=1]{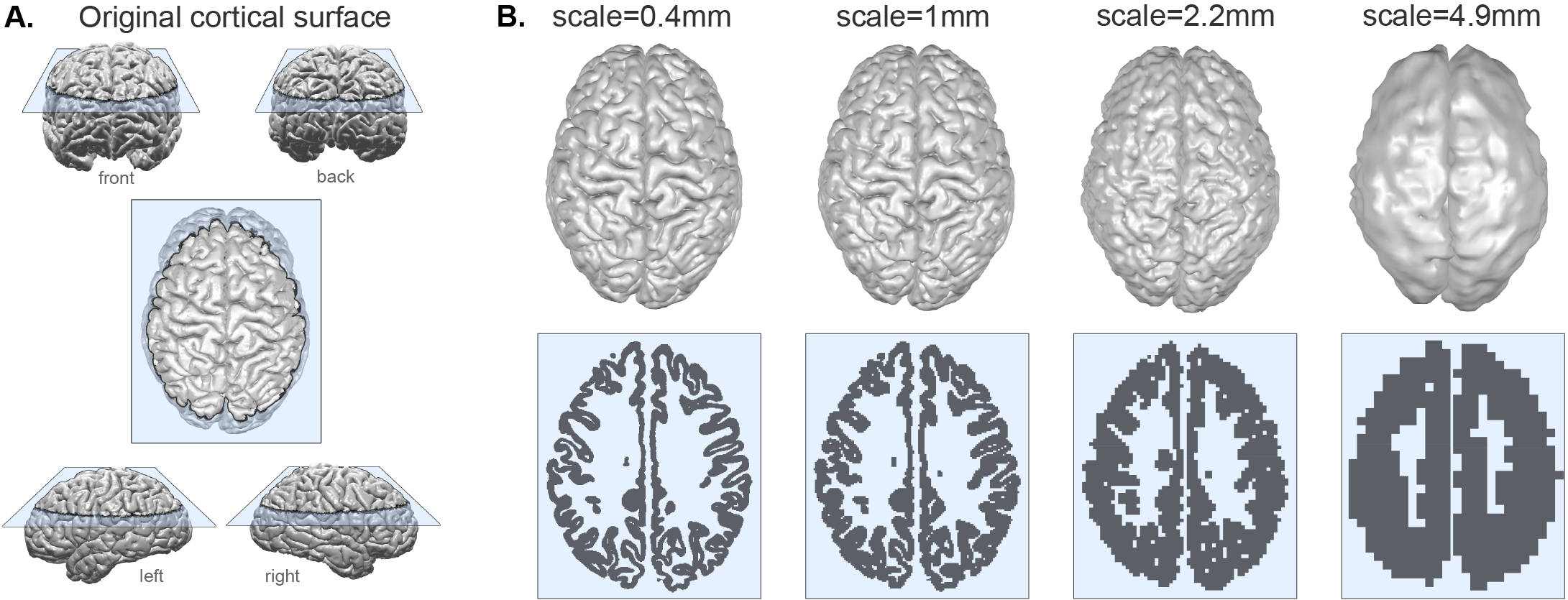}
\caption{Coarse-graining a cortex at different scales. \textbf{A}: Example original pial surface from a healthy human viewed from varying angles. \textbf{B}: Same example cortical surface from A, coarse-grained to different spatial scales. Top row shows resulting pial surfaces (with a small amount of smoothing applied for visualisation purposes here). Bottom row shows the corresponding voxelisation at each scale through the slice indicated by blue plane in panel A. Note that the actual size of the brains for analysis are rescaled (see Methods and Fig.~\ref{fig:scaling}); we display all brains scaled at an equal size here for the ease of visualisation of the method.}
\label{fig:meltingprocess}
\end{figure}

Concretely, our proposed method requires the bounding pial and white matter surfaces of the cortical ribbon as input. We obtained these surfaces based on reconstructions from magnetic resonance imaging data in 11 different primate species. Algorithmically, we then segment the space between the original pial and white matter surfaces into a 3D grid of boxes of the desired scale $\lambda$, where each box is a cube of dimensions $\lambda \times \lambda \times \lambda$. We also term the 3D grid of cubes ``voxelisation'', as it effectively captures the cerebral cortex as voxels in 3D space (Fig.~\ref{fig:meltingprocess}B bottom row). At any given scale, or voxel size, this process effectively erases morphological features (folds) that are smaller than the cube size. Visually, increasing the voxel size appears as if the cortex is ``melting'' and ``thickening'' (Fig.~\ref{fig:meltingprocess}, and videos: \url{https://bit.ly/3CDoqZQ}).

A more technical and detailed description and discussion of the algorithm is provided in Suppl.~\ref{coarsegrainingalgo}. Note this method has also no direct dependency on the original MR image resolution, as the inputs are \textit{smooth} grey and white matter surface meshes reconstructed from the images using strong (bio-)physical assumptions and therefore containing more fine-grained spatial information than the raw images (see also Suppl.~\ref{scan_res}).

\subsection{Rescaling coarse-grained outputs for analysis}
Morphological properties, such as cortical thicknesses measured in our ``melted'' brains are to be understood as a thickness relative to the size of the brain. Therefore, to analyse the scaling behaviour of the different coarse-grained realisations of the same brain, we apply an isometric rescaling process that leaves all dimensionless shape properties unaffected (more details in Suppl.~\ref{obtaining_scaling_law}). Conceptually, this process fixes the voxel size, and instead resizes the surfaces relative to the voxel size, which ensures that we can compare the coarse-grained realisations to the original cortices, and test if the former, like the latter, also scale according to Eqn. (\ref{eqn1_scalinglaw}). Resizing, or more precisely, shrinking the cortical surface is mathematically equivalent to increasing the box size in our coarse-graining method. Both achieved an erasure of folding details below a certain threshold. After rescaling, as an example, the cortical thickness also shrinks with increasing levels of coarse-graining, and never exceeds the thickness measured at native scale. 

\subsection{Independent morphological measures of shape}
To better characterise the coarse-grained cortices in terms of their similarity in offset, we use a previously introduced  \citep{wang2021} set of independent measures, $K$, $I$ and $S$, that summarise the morphometry of the cortex in a natural and statistically robust way. In this framework, isometrically scaled copies of the same morphometry all map onto a line along the $I=\log A_t + \log A_e + \log T^2$ direction, which is perpendicular to a $K \times S$ plane that fully summarises their shape. $K=\log A_t - \frac{5}{4}\log A_e +\frac{1}{4}\log T^2$ is the direction defined by the offset $k$ of the scaling law Eqn.~(\ref{eqn1_scalinglaw}), while direction $S =\frac{3}{2}\log A_t + \frac{3}{4}\log A_e -\frac{9}{4}\log T^2$ captures the remaining information about shape, and can be regarded as a simple measure of morphological complexity. Ref \citep{wang2021} provides a detailed derivation and demonstration of the superior sensitivity and specificity of these new morphometric measures. The advantage of using this framework here is that we can assess the offset $K$ (and shape term $S$) without interference by isometric size effects, including any re-scaling procedures.

\subsection{Data and processing}

With the exception of the marmoset data, all other cortical surface reconstructions were based on healthy individual brains.

\subsubsection{Human data}
To study healthy human adults, we used the Human Connectome Project (HCP) MRI data, available at \url{https://db.humanconnectome.org/}  \citep{van_essen2012}, obtained using a 3T Siemens Skyra scanner with 0.7~mm isotropic voxel size. We used the HCP minimally pre-processed FreeSurfer data output, which provided the pial and white matter surface meshes we required. We selected five random subjects in the age category 22-25 y.o. and show one example subject (103414) in the main text, and the remaining subjects in Suppl. \ref{suppl_species_detail}.

To study the alterations associated with human ageing, we used T1 and T2 weighted MRI brain scans from The Cambridge Centre for Ageing and Neuroscience (Cam-CAN) dataset (available at \url{http://www.mrc-cbu.cam.ac.uk/datasets/camcan/}  \citep{shafto2014cambridge,taylor2017cambridge}). Cam-CAN used a 3T Siemens TIM Trio System with 1~mm isotropic voxel size (for more details see  \citep{shafto2014cambridge,taylor2017cambridge}). 
From the Cam-CAN dataset we retained 644 subjects that successfully completed preprocessing (with Freesurfer recon-all) without errors. From these subjects we selected all subjects between the ages of 17-25 inclusive (forming the 20 y.o. cohort, n=27); we also selected all subjects between the ages of 77-85 inclusive (forming the 80 y.o. cohort, n=86).

To confirm the ageing results, we also obtained an independent dataset from the Nathan Kline Institute (NKI)/Rockland sample \citep{Nooner2012} (\url{http://fcon_1000.projects.nitrc.org/indi/pro/nki.html}) using the same procedure as described for the CamCAN dataset.

The MR images of both CamCAN and NKI datasets were first preprocessed by the FreeSurfer 6.0 pipeline \textit{recon-all}, which extracts the grey-white matter boundary as well as the pial surface. These boundaries were then quality checked by visual inspection for particularly the young and old cohorts and manually corrected where needed.

For all three datasets, we obtained the pial and white matter surfaces for further analysis. In the current work, the analysis is always hemisphere based, as in our previous work \citep{science2015,pnas2016}. We did not perform a more regionalised analysis, which is also possible  \citep{commbiol2019,leiberg2021}. Future work using the principle demonstrated here can be directly extended to derive regionalised measures across scales.

\subsubsection{Non-human primate data}
\textbf{Macaque}\\
Rhesus Macaque MRI scans were carried out at the Newcastle University Comparative Biology Centre. Macaques were trained to be scanned while awake and sat in a primate chair. Both T1 weighted MP-RAGE and T2 weighted RARE sequences were acquired, using a vertical MRI scanner (Biospec 4.7 Tesla, Bruker Biospin, Ettlingen, Germany).

Scans were processed using a custom macaque MRI pipeline, incorporating ANTs, SPM, FreeSurfer and FSL. Briefly, this involved the creation of precursor mask in SPM, denoising (ANTs DenoiseImage) and debiasing (ANTs N4BiasFieldCorrection, and Human Connectome BiasFieldCorrection script), creation of a final mask in SPM and then processing in FreeSurfer (using a modified version of the standard FreeSurfer processing pipeline). Reconstructed pial and white matter surfaces were visually quality controlled in conjunction with the MR images.

\textbf{Marmoset}\\
The marmoset MRI structural scan was collected as part of the development of the NIH marmoset brain atlas  \citep{LiuEtal2018,LiuEtal2020}. Data was collected ex vivo from a 4.5-year-old male marmoset using a T2-star weighted 3D FLASH sequence using a horizontal MRI scanner (Biospec 7 Tesla, Bruker Biospin, Ettlingen, Germany).

A total of ten scans were collected and averaged into one final image. In combination with scans from other modalities, cortical boundaries were manually delineated on each coronal slice. Boundaries were then refined through comparisons with other atlases.
Volumetric data was then converted to surfaces using a custom pipeline involving an intermediate generation of high-resolution mesh data  \citep{Madan2015}, decimation  \citep{Madan2016}, and remeshing.
Reconstructed pial and white matter surfaces were visually quality controlled in conjunction with the MR images.

\textbf{Other non-human primates (NHPs)}\\
The remaining NHP MRIs and subsequent brain surface extraction is detailed in  \citep{ardesch2022,bryant2021}, and provided to the authors in a processed format. Briefly, a range of specialised scanners were used to acquire optimal images for each species. FreeSurfer 6.0 with some modifications was used for surface reconstruction, complemented by FSL, ANTS, and Matlab. All surfaces were visually inspected for accuracy and consistency across datasets.

\subsubsection{Comparative neuroanatomy data}
The comparative neuroanatomy dataset for different mammalian species is the same as previously published  \citep{science2015}. Note that for this dataset, we only had numerical values for the total and exposed surface area, as well a average cortical thickness estimates. We did not perform any analysis across scales in this dataset (hence surfaces were not required), but only used it as a reference dataset.

\subsection{Statistical analyses}

Briefly, linear regression is used in either a mixed-effect model to capture effects across individuals and species, or in simple fixed-effect settings to estimate regression slopes to obtain fractal dimension.

In the final part of Results, we analyse the effect between a group of 20-year-olds and 80-year-olds. Effect size is calculated as Cohen's D between the two groups.

Throughout the paper, statistical significance is not a crucial argument, and we report p-values only for reference and completeness.

More details can be seen in the analysis code, where the reader can directly reproduce all main result figures.

\subsection{Code and data availability}
The code for coarse-graining has been integrated into our MATLAB toolbox Cortical Folding Analysis Tools: \url{https://github.com/cnnp-lab/CorticalFoldingAnalysisTools}, include a graphical user interface. Users will also see the latest updates in this repository.

The analysis code underpinning this paper is published on GitHub: 
\url{https://github.com/cnnp-lab/2024_Folding_scales/} 

The post-processing data (i.e. ``voxelisations'' and derived metrics) are uploaded on Zenodo: \url{DOI:10.5281/zenodo.12820611}

The data, together with the code, will allow readers to reproduce of our main results.

\section{Results\label{results}}

\subsection{All primate brains follow the same scaling law across spatial scales\label{results1}}

\begin{figure}[h!]
\hspace{-0.5cm}
\includegraphics[scale=1]{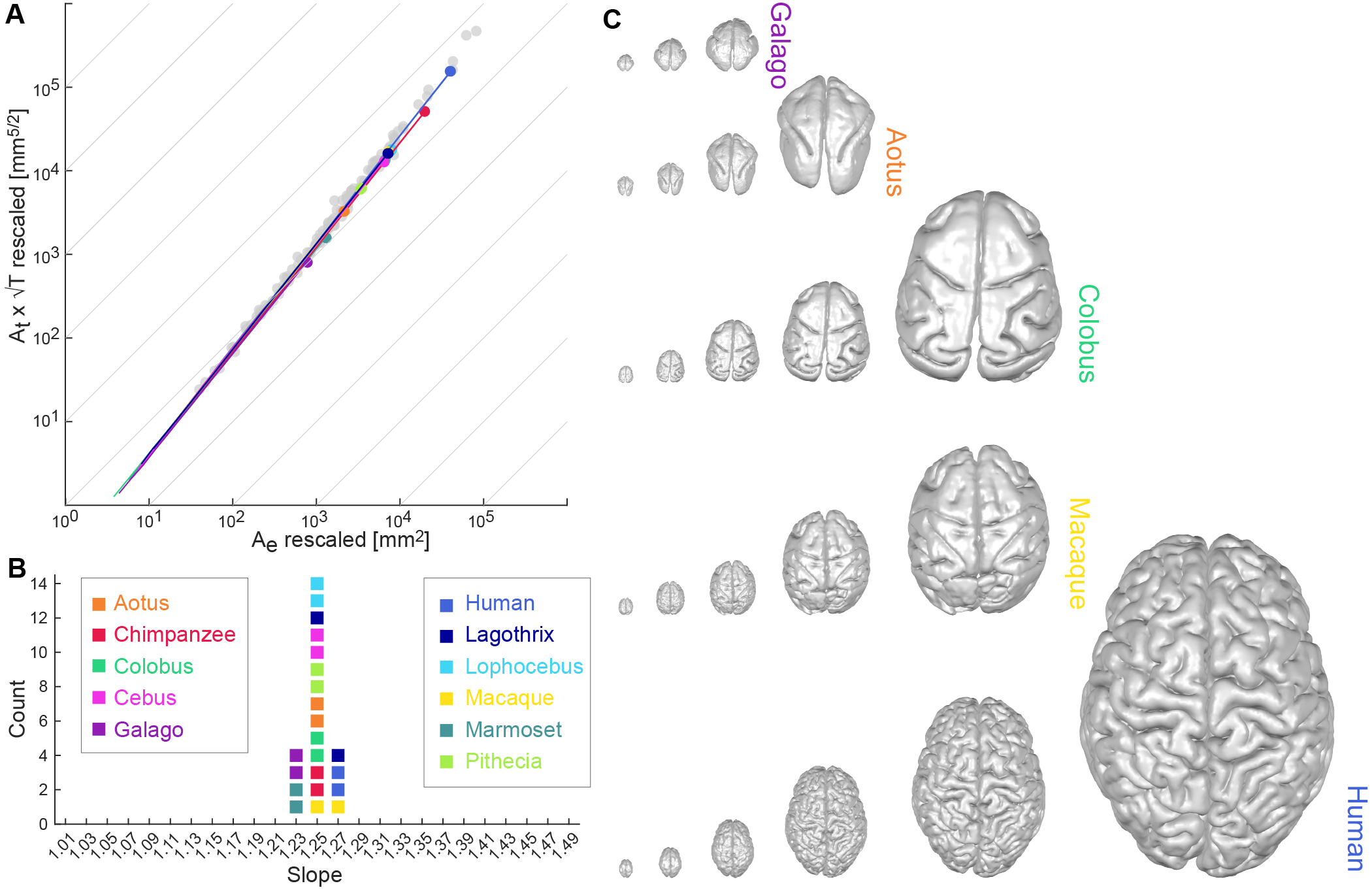}
\caption{\textbf{Universal scaling law for 11 coarse-grained primate brains.} \textbf{A:} Coarse-grained primate brains are shown in terms of their relationship between log10($A_t \times \sqrt{T}$) \textit{vs.} log10($A_e$). Each solid line indicates a cortical hemisphere from a primate species. Thin grey lines indicate a slope of 1 for reference. Filled circles mark the data points of the original cortical surfaces. Grey data points are plotted for reference and show the comparative neuroanatomy dataset across a range of mammalian brains \citep{science2015}. \textbf{B:}~Slopes ($\alpha$) of the regression of data points in A for each species. For each species, two data points are shown, one per hemisphere. Colour-code for each species is maintained throughout the whole figure. \textbf{C:} Rescaled brain surfaces visualised for five example species at different levels of coarse-graining. }
\label{fig:scaling}
\end{figure}

We have analysed cortices of 11 different primate species: aotus, cebus, chimpanzee, colobus, galago, human, lagothrix, lophocebus, macaque, marmoset, and pithecia, and various cohorts of human subjects. We applied our coarse-graining procedure to their pial and white matter surfaces, and empirically determined (i) that all species followed a power law (linear regression $R^2>0.999$ for all species); (ii) the slope of said power law is $\alpha=1.255$ on a group level (CI: $[1.254 \; 1.256]$) using linear mixed effect modelling (see Fig.~\ref{fig:scaling} for visualisation, and Suppl. \ref{suppl_species_detail} for a detailed breakdown by species); and importantly, (iii) all species also show a similar offset $\log k \approx -0.65263$, with a standard deviation of intercept across species estimated at $0.02$ from linear mixed effect modelling in $\log k$. 

Taken separately, the scaling for each species is proof that their cortices are self-similar with the same scaling: they each approximate a fractal with fractal dimension $d_f = 2.5$. Considering all species together, different species also overlap substantially (similar offset), and only differ from each other in the range of scales over which the approximation is valid (see Suppl.~\ref{species-specific} and \ref{morpho}). Thus, as Fig.~\ref{fig:scaling} illustrates, the data supports a \textit{universal} scaling law across primate species and spatial scales:

\begin{equation}
    A_t(\lambda) T(\lambda)^{\frac{1}{2}} = k  A_e(\lambda) ^{\frac{5}{4}},
    \label{eq:scaling}
\end{equation}
with $k = 0.2277.$

%As a negative control, we also applied our voxelisation procedure to a walnut shape (see Suppl. \ref{validation_nonbrain}). While the shape of the brain has been likened to a walnut before, we observe a distinctly lower slopes (walnut 1: $\alpha=1.191$, $CI=[1.185  \;  1.196]$; walnut 2: $\alpha=1.193$, $CI=[1.187  \;  1.198]$) than the primate brains, and a varying offset. This result demonstrates that the scaling (\ref{eq:scaling}) is indeed a characteristic of the primate brains we studied, and is not an artifact of the algorithm. 

\subsection{Primate brains at different spatial scales are morphometrically similar to each other and other mammalian species\label{results2}}

To better characterise the coarse-grained cortices in terms of their similarity in offset, we use a set of independent morphometric measures, $K$, $I$ and $S$, that summarise the morphometry of the cortex in a natural and statistically robust way. We can therefore assess the offset $K$ (and shape term $S$) without interference by the isometric size or rescaling.

\begin{figure}[h!]
%\hspace{-3.2cm}
\centering
\includegraphics[scale=1]{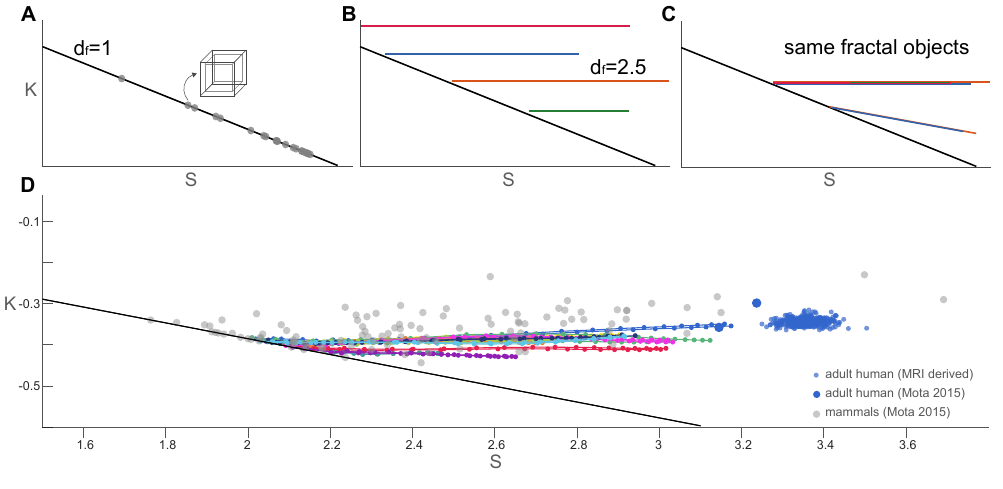}
\caption{\textbf{Trajectories of coarse-grained primate cortices and other mammalian and human brains in $K \times S$ plane.} \textbf{A:} Straight trajectories indicate self-similarity (described by a scaling law). In particular, the black line here indicates objects with $A_e=A_t$ for all scales, such as the box of finite thickness with a fractal dimension $d_f=1$ (grey data points). This line is reproduced in all subpanels for reference. \textbf{B:} Morphological trajectories of multiple hypothetical fractal objects are shown which. A flat trajectory (constant $K$) correspond to $d_f=2.5$ in this space. However, these objects are clearly different fractals with different values of $K$.  \textbf{C:} Hypothetical objects with overlapping straight morphological trajectories indicate multiple realisations of the same fractal object. Flat trajectories (constant $K$) correspond to $d_f=2.5$. The two hypothetical objects with a decreasing $K(S)$ correspond to $2<d_f<2.5$ \textbf{D:} Projecting our actual data into the normalised $K \times S$ plane showing the coarse-grained primate brains (same as in Fig.~\ref{fig:scaling}) as data point connected with solid lines (colour-code same as Fig.~\ref{fig:scaling}). Different mammalian brains are shown as grey scatter points, and adult human data points are blue.}
\label{fig:KxS}
\end{figure}

We can measure $K$ and $S$ for any object, but a fuller expression is captured by the \emph{trajectory} of said object as a function of coarse-graining in the $K \times S$ plane. This is a very convenient and informative way of summarising an object: self-similar objects correspond to straight trajectories as the $K \times S$ plane is in log-log space. In particular, objects without any folds or protrusion (i.e., convex, such as the box with finite thickness in Fig.~\ref{fig:KxS}~A) correspond to the line $K=-\frac{1}{9}S$, as $A_e = A_t$ for all levels of coarse-graining. Horizontal trajectories (constant $K$) represent fractal objects with fractal dimension $d_f=2.5$ (Fig.~\ref{fig:KxS}~B). And finally, in the $K \times S$ plane, a group of objects can said to be ``universal'' when their trajectories overlap, so that they can all be regarded as coarse-grained versions of one another (Fig.~\ref{fig:KxS}~C). 

%In the following, we focus on the $K \times S$ plane to better quantify self-similar scaling, fractal dimension, and universal fractal shape of coarse-grained primate brains. Note the $K \times S$ plane is more informative and intuitive, as objects following a scaling law will form straight lines over coarse-graining, independent of isometric size. More specifically, simple, non-fractal objects will lie on a single line described by $K=-\frac{1}{9}S$ (where $A_e$ is always the same as $A_t$ regardless of scale), such as a box with finite thickness (Fig.~\ref{fig:KxS} a). Furthermore, fractal objects with a fractal dimension of 2.5 will form flat lines with coarse-graining in this space (slope=0, but possibly different offsets, see Fig.~\ref{fig:KxS} b). Finally, if flat lines overlap, it would indicate a single universal fractal shape with fractal dimension of 2.5 (slope=0, same offset, see Fig.~\ref{fig:KxS} c).

Primate cortices (Fig.~\ref{fig:KxS}~D) display a nearly invariant $K$ in all cases. But, over all levels of coarse-graining, $K$ also remains near-invariant in all \textit{trajectories} as $S$ decreases, resulting in a set of horizontal lines that largely overlap with each other and other mammalian species. The variance in $K$ across scales and all 11 species is $<0.01$, which is at least an order of magnitude lower than the variance in $S$. Primate brains therefore have all three characteristics of self-similarity, fractality (with $d_f=2.5$), and universality (invariant $K$ for all scales and species) at the same time.

%Fascinatingly, Fig.~\ref{fig:KxS}~d shows normalised $K$ and $S$ values for each primate species across scales, and demonstrates that the coarse-graining procedure generates horizontal trajectories that largely overlap with each other and other mammalian species. While $S$ decreases with increasing scale $\lambda$, $K$ remains nearly constant for all primate brains. 

Thus, coarse-grained primate cortices are morphometrically similar to, and in terms of the universal law, `as valid as' actual existing mammalian cortices\footnote{Note, of course, that the coarse-grained brain surfaces are an output of our algorithm alone and not to be directly/naively likened to actual brain surfaces, e.g. in terms of the location or shape of the folds. Our comparisons here between coarse-grained brains and actual brains is purely on the level of morphometrics across the whole cortex.}. In contrast, we tested various non-brain objects, and while e.g. the walnut, and bell pepper form (partially) straight lines, they vary in both K and S (see Suppl.~\ref{validation_nonbrain}). These objects may have a fractal regime, but their fractal dimension is not 2.5, nor are they similar to primate or mammalian brains in terms of $K$ or $S$. Furthermore, supplementary~\ref{validation_jitter} underscores the algorithmic and statistical robustness of these results using multiple realisations of the coarse-graining procedure on the same object. %Indeed, visually and qualitatively, appropriately-scaled primate cortices with the same value of $S$ look very similar to one another, as can be seen in the example pial surfaces at the top of Fig.~\ref{fig:KxS}. 
In this framework, our main result can thus be expressed simply: for all the cortices we analysed, and for none of the non-cortices, coarse-graining will leave $K$ largely unaffected, while morphological complexity $S$ will decrease.

\subsection{Morphometric measures as functions of scale reveal scale-specific effects of ageing\label{results3}}
%In the final part of our work, we show how our algorithm and the associated new understanding of brain morphology may become useful in applications. As an example, we will focus on how the ageing process affects human cortical morphology across scales. In Fig.~\ref{fig:ageing}~A, we compare $K(\lambda)$ as a function of scale $\lambda$ for a young (20 year old) \textit{vs.} an old (80 year old) group of human brains. Strikingly, the difference in $K$ between the groups takes a U shape, where the ageing effect peaks at about 2mm (cp. Fig.~\ref{fig:ageing}~C middle column). For scales over 5mm, the aging process appears to have little effect on the larger cortical morphological features (cp. Fig.~\ref{fig:ageing}~C right column). Corresponding effect plots are shown for $A_t$ (Fig.~\ref{fig:ageing}~B) for reference. Suppl.~\ref{ageing} gives a more detailed description, validation and discussion of these results. In summary, the ageing process displays a scale-specific effect, or expressed in other words, morphological biomarkers of ageing may be best detected at a spatial scale of \~2mm.

\begin{figure}[h!]
\begin{adjustwidth}{-1cm}{0cm}
% \centering
\includegraphics[scale=1]{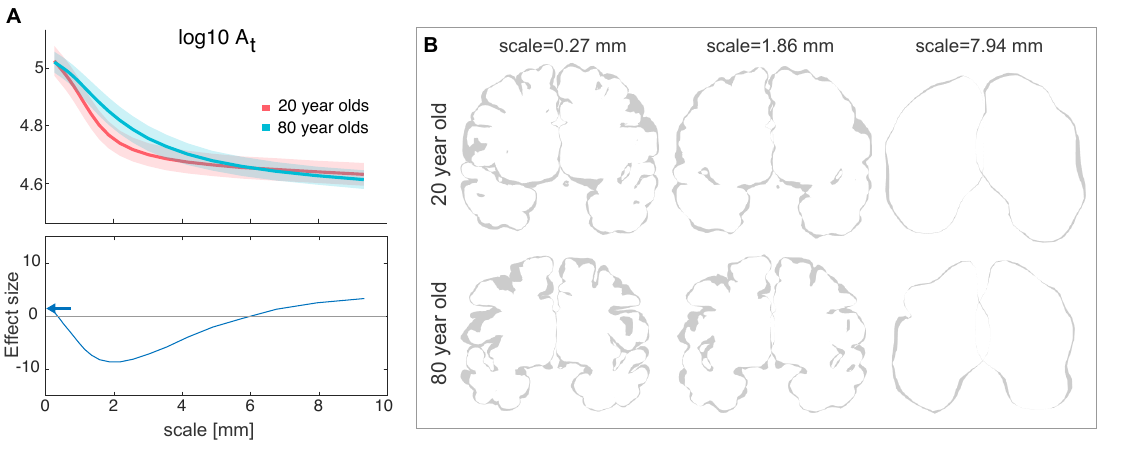}
\caption{\textbf{Human ageing shows differential effects depending on spatial scale.} \textbf{A:} Top: $A_t(\lambda)$ is shown for a group of 20-year-olds (red, n=27) and a group of 80-year-olds (blue, n=86). Mean and standard deviation are shown as the solid line and the shaded area respectively. Bottom: Effect size (measured as ranksum z-values) between the older and younger groups at each scale. Positive effect indicates a larger value for the younger group. Blue arrows indicate the effect size at ``native scale'' i.e. using the original freesurfer grey and white matter meshes. \textbf{B:} Coronal slices of the pial surface of an example 20-year-old subject and an 80-year-old subject at different scales (columns).}
\label{fig:ageing}
\end{adjustwidth}
\end{figure}

In the final part of our work, we show how our algorithm and the associated new understanding of brain morphology may become useful in applications. As an example, we will focus on how the ageing process affects human cortical morphology across scales. In Fig.~\ref{fig:ageing}~A, we compare the total surface area $A_t(\lambda)$ as a function of scale $\lambda$ for a young (20 year old) \textit{vs.} an old (80 year old) group of human brains. The difference in $A_t$ between the groups takes a U shape, and the strongest effect is seen at approximately 2 millimeter (greatest effect size of -8.635 seen at scale 2.188~mm), where older subjects have higher $A_t$. For scales over $\sim 5$~mm, and under $\sim 0.5$~mm the differences become relatively small, suggesting the ageing process has less effect on the largest and smallest cortical morphological features. Finally, we reproduced these results in an independent dataset in Suppl.~\ref{ageing_nki}.

In this particular example, the scale-dependency of morphological measures can be visually and intuitively understood by looking at the reconstructed surfaces at each scale: Fig.~\ref{fig:ageing}~B shows some coronal slices of the cortical surface. At scale 0.27~mm, the gyri in the younger subjects are densely packed, but the older subjects show the expected widening between gyral walls and decrease in gyral surface area at the crown (see e.g.  \citep{Jin2018,Madan2019} for recent investigations and references therein). At 1.86~mm, the younger cortices have already partially ``melted'', erasing most small sulci between the densely packed gyri. In the older humans, however, the gyri are less dense, the sulci more open, and thus, at 1.86~mm, most gyri and sulci have not been erased yet.  At scale 7.94~mm, both young and old brains have ``melted'' down to similarly near-lissencephalic cortices. For a more detailed multiscale investigation over the entire human lifespan, please refer to our new preprint \cite{leiberg2024multiscale}

More broadly, one can regard the melting process as a way of determining how cortical area is allocated across different scales. As the cortex `melts', the contributions to the total area from features smaller than the cut-off scale are eliminated. In the example in Fig.~\ref{fig:ageing}, for instance, we can say about half ($10^{5-4.7}=10^{0.3} \approx 2$) the total area in the 80 y.o. cortices is present in features smaller than 4~mm.

\section{Discussion\label{discussion}}

We have devised a new way of expressing the morphology of the mammalian cerebral cortex, as the flow in the values of morphometric measures over a range of spatial scales. This was achieved by coarse-graining cortical surfaces, erasing morphological features smaller than the specified scale, while preserving surface integrity. After applying this method to the cortices of 11 primate species, we have shown that all these diverse cortices are approximations of \textit{the same archetypal self-similar (fractal) shape}. Most of their morphological diversity can be ascribed to the species-specific ranges of spatial scales over which each approximation is valid, with the smallest scale being an invariant multiple of cortical thickness. This was a proof-by-construction of fractality, a step beyond the usual box-counting approach, which also yields scale-dependent morphometrics. As a proof-of-principle we showed that healthy human ageing has highly scale-dependent effects in a range of morphometrics.

\subsection{Advantages and advances}
Compared to previous literature, we can summarise our main contribution and advance as follows:
(i) We are showing for the first time that representative primate species follow the exact same fractal scaling – as opposed to previous work showing that they have a similar fractal dimension \cite{Hofman1985,Hofman1991}, i.e. slope, but not necessarily the same offset, as previous methods had no consistent way of comparing offsets. 
(ii) Previous work could also not show direct agreement in morphometrics between the coarse-grained brains of primate species and other non-primate mammalian species.
(iii) Demonstrating in proof-of-principle that multiscale morphometrics, in practice, can have much larger effect sizes for classification applications. This moves beyond our previous work where we only showed the scaling law across \cite{mota_herculano-houzel_2014} and within species \cite{pnas2016}, but all on one (native) scale with comparable effect sizes for classification applications \cite{pnas2016}. 

In simple terms: we know that objects can have the same fractal dimension, but differ greatly in a range of other shape properties. However, we demonstrate here, that representative primate brains and mammalian brain indeed share a range of other key shape properties, on top of agreeing in fractal dimension. This suggests a universal blueprint for mammalian brain shape and a common set of mechanisms governing cortical folding. As a practical additional outcome of our study, we could show that our novel method of deriving multiscale metrics can differentiate subtle morphological changes much better (4 times the effect size) than the metrics we have been using so far at a single native scale.

Expressing cortical morphology as a function of scale is more detailed than a list of summary morphometric measures, and more informative than the mere listing of every sulcus and gyrus. We propose this new syntax as the basis for a more rigorous characterisation of brain morphology and morphological changes. A clear advantage is that some biological processes may only act on a specific spatial scale, leaving other scales untouched (ageing in our example). By disambiguation of the spatial scale, it allows for an extra dimension of understanding, and tracking of biological processes. In the future, this approach can also be extended to cortical development and to various degenerative \citep{pnas2016} and congenital \citep{wang2021} neuropathic conditions, especially if combined with a regionalised version of this method applied to specific cortical regions~ \citep{commbiol2019} or local patches \citep{leiberg2021}. 

\subsection{Implications of universality}
Empirically, the main result of this paper is the demonstration of a universal self-similar scaling (Eqn.~\ref{eq:scaling}) for primate, and presumably mammalian cortices. There are two aspects of this universality: first that for each and every cortex the value of $K$ \textit{remains} the same for all scales as one removes substructures smaller than a varying length scale (or equivalently, that the fractal dimension is almost exactly $2.5$ in all cases). Second, that the value of $K$ for the cortices of different species is approximately the same, as previously observed (Eqn.~\ref{eqn1_scalinglaw}) across species \citep{science2015} and individuals \citep{pnas2016}. One could imagine a set of objects for which one but not the other aspect of the universality in $K$ holds true. However, the fact that both universalities hold true is significant. It suggests the existence of a single highly conserved mechanism for cortical folding, operating on all length scales self-similarly with only a few morphological degrees of freedom. It also hints at the possibility of deriving cortical scaling from some variational principle. Finally, this dual universality is also a more stringent test for existing and future models of cortical gyrification mechanisms at relevant scales, and one that moreover is applicable to individual cortices. For example, any models that explicitly simulate a cortical surface as an output could be directly coarse-grained with our method and the morphological trajectories can be compared with those of actual human and primate cortices. The simulated cortices would only be `valid' in terms of the dual universality, if it also produces the same morphological trajectories.\footnote{Note, we do not suggest to directly compare coarse-grained brain surfaces with actual biological brain surfaces. As we noted earlier, the coarse-grained brain surfaces are an output of our algorithm alone and are not to be directly/naively likened to actual brain surfaces, e.g. in terms of the location or shape of the folds. Our comparisons here between coarse-grained brains and actual brains is purely on the level of morphometrics across the whole cortex.}

The scaling itself does not imply or favour any particular proposed gyrification model (ours \citep{science2015} included), and all results in this paper are agnostic about this choice. Indeed, our previously proposed model \citep{science2015} for cortical gyrification is very simple, assuming only a self-avoiding cortex of finite thickness experiencing pressures (e.g. exerted by white matter pulling, or by CSF pressure). The offset $K$, or `tension term', precisely relates to these pressures, leading us to speculate that subtle changes in $K$ correlate with changes in white matter property \cite{pnas2016,wang2021}. In the same vein of speculation, the scale-dependence of $K$ shown in this work might therefore be related to different types of white matter that span different length scales, such as superficial \textit{vs.} deep white matter, or U-fibres \textit{vs.} major tracts. However, there are also challenges to the axonal tension hypothesis \cite{xu_axons_2010}. Indeed, white matter tension differentials in the developed brain may not explain location of folds, but instead white matter tension may contribute to a whole-brain scale `pressure' during development that drives the folding process overall. Aside from speculations about the biological interpretation, the simplicity of the highlighted scaling law parallels many complex phenomena in nature that display simple and universal scaling that can be derived from first principles  \citep{barenblatt1996scaling,West1997,gagler_scaling_2022}. In addition, recent results suggest simplicity and symmetry are generically favoured on statistical-ensemble grounds by evolution  \citep{Johnston2022}. Our model correctly predicts the scaling law (Eqn.~\ref{eqn1_scalinglaw}), but a more complete explanation for cortical gyrification is probably far more complex \citep{Quezada2020} than can be accounted by such a simple model. 

One specific example of said complexity are the exact patterns, locations, depth, and features of gyri and sulci. We know such patterns to be, for example, variable but also somewhat heritable in humans, whilst in macaques such patterns are relatively preserved across the species. Our work does not explain any of these observations, nor are the coarse-grained versions of human brains supposed to exactly resemble the location/pattern/features of gyri and sulci of other primates. The similarity we highlighted here are on the level of summary metrics, and our goal was to highlight the universality in such metrics to point towards highly conserved quantities and mechanisms.

\subsection{Biological plausibility and implications}

The observation that with increasing voxel sizes, the coarse-grained cortices tend to be smoother and thicker is particularly interesting: the scaling law in Eq.~\ref{eq:scaling} can be understood as thicker cortices ($T$) form larger folds (or are smoother i.e. less surface area $A_t$) when brain size is kept constant ($A_e$). This way of understanding has also been vividly illustrated by using the analogy of forming paper balls with papers of varying thickness in Mota \textit{et al.}\cite{mota_herculano-houzel_2014}: to achieve the same size of a paper ball ($A_e$), the one that uses thicker paper ($T$) will show larger folds (or is smoother i.e. less surface area $A_t$) than the one using thinner paper.\footnote{Note that these observations should not be interpreted in absolute terms, smoother brains are of course not always thicker (measured in millimeters) than folded brains, neither in the real-world, nor in our outputs. These statements are formed under the assumption of the same isometric size of the cortex, or the paper ball.} The scaling law can therefore be understood as a physically and biologically plausible statement and our algorithm yields results in line with the scaling law. 

More broadly, the interaction between brain development and evolution may also benefit from a scale-specific understanding. This may be important in elucidating what in cortical morphology is selected for by evolution, what is determined by physics; what is specified by genes, and what is emergent. For example, one can estimate the number of structural features at each scale ($A_t(\lambda)$ as multiples of $A_0(\lambda)$), and it will be interesting to correlate this number to other quantifiers of cortical structure, such as number of neurons \citep{mota_herculano-houzel_2014}, number of functional areas \citep{molnar2014} or the number of cortical columns \citep{Kaas2012}, possibly over different stages of development. Generally, the larger the cortical feature (i.e. from gyri to functional areas to lobes to hemispheres), the earlier during development it appears \citep{ZILLES2013,Garcia2018}, and more broadly it is conserved over kinship~ \citep{Pizzagalli2020} and phylogeny~ \citep{HEUER2019,Valk2020}. It thus seems likely that comparative neuroanatomical methods \citep{Mars2014,Croxson2017} may be directly used to identify and contrast structures in coarse-grained cortices of more highly gyrified species with their analogues in less-gyrified species. One could then perhaps specify when evolution conserves and when it invents old and new cortical features.

From an application perspective, our final result illustrates clearly that the surface area difference between older and younger subjects at ``native'' scale (i.e. original freesurfer surfaces) is negligible (effect size smaller than 2 standard deviations). However, in our analysis across scales, there is a clear optimal scale at $\sim 2$~mm where the effect size is maximised between older and younger subjects (effect size is -8 standard deviations). For most classification applications in biology and medicine, the increased effect size and hence separability of groups in the scale-dependent morphometrics represent a huge advance over the native scale.

\subsection{Outlook}
Our work here was limited to summary descriptors of entire cortical hemispheres, but future work will explore extensions of these methods to lobes and cortical areas, similarly to  \citep{commbiol2019,leiberg2021}. This will generate precise characterisations of the morphological differences between phylae and across developmental stages, and perhaps pinpoint the time and location of morphological changes leading to congenital and neurodegenerative conditions. Ultimately, we hope this new framework for expressing and analysing cortical morphology, besides revealing a hitherto hidden regularity of nature, can become a powerful tool to characterise and compare cortices of different species and individuals, across development and ageing, and across health and disease.

\FloatBarrier
\newpage
\nolinenumbers
\bibliography{references}

\begin{thebibliography}{59}
\providecommand{\natexlab}[1]{#1}
\providecommand{\url}[1]{\texttt{#1}}
\expandafter\ifx\csname urlstyle\endcsname\relax
  \providecommand{\doi}[1]{doi: #1}\else
  \providecommand{\doi}{doi: \begingroup \urlstyle{rm}\Url}\fi

\bibitem[Zhang and Sejnowski(2000)]{Zhang2000}
Kechen Zhang and Terrence~J. Sejnowski.
\newblock A universal scaling law between gray matter and white matter of
  cerebral cortex.
\newblock \emph{Proceedings of the National Academy of Sciences}, 97\penalty0
  (10):\penalty0 5621--5626, 2000.
\newblock \doi{10.1073/pnas.090504197}.
\newblock URL \url{https://www.pnas.org/doi/abs/10.1073/pnas.090504197}.

\bibitem[Francis et~al.(2009)Francis, Frederic, Sylvie, Prasanna, and
  Henri]{Francis2009}
Cassot Francis, Lauwers Frederic, Lorthois Sylvie, Puwanarajah Prasanna, and
  Duvernoy Henri.
\newblock Scaling laws for branching vessels of human cerebral cortex.
\newblock \emph{Microcirculation}, 16\penalty0 (4):\penalty0 331–344, 2009.
\newblock \doi{10.1080/10739680802662607}.

\bibitem[Karbowski(2011)]{Karbowski2011}
Jan Karbowski.
\newblock Scaling of brain metabolism and blood flow in relation to capillary
  and neural scaling.
\newblock \emph{PLoS ONE}, 6\penalty0 (10), 2011.
\newblock \doi{10.1371/journal.pone.0026709}.

\bibitem[Mota and Herculano-Houzel(2014)]{mota_herculano-houzel_2014}
Bruno Mota and Suzana Herculano-Houzel.
\newblock All brains are made of this: A fundamental building block of brain
  matter with matching neuronal and glial masses.
\newblock \emph{Frontiers in Neuroanatomy}, 8, 2014.
\newblock \doi{10.3389/fnana.2014.00127}.

\bibitem[Mota and Herculano-Houzel(2015)]{science2015}
Bruno Mota and Suzana Herculano-Houzel.
\newblock Cortical folding scales universally with surface area and thickness,
  not number of neurons.
\newblock \emph{Science 349(6243):74–77}, 2015.
\newblock \doi{10.1126/science.aaa9101}.
\newblock URL \url{https://doi.org/10.1126/science.aaa9101}.

\bibitem[Wang et~al.(2016)Wang, Necus, Kaiser, and Mota]{pnas2016}
Yujiang Wang, Joe Necus, Marcus Kaiser, and Bruno Mota.
\newblock Universality in human cortical folding in health and disease.
\newblock \emph{PNAS}, 2016.
\newblock \doi{10.1073/pnas.1610175113}.
\newblock URL \url{https://doi.org/10.1073/pnas.1610175113}.

\bibitem[Wang et~al.(2019)Wang, Necus, Rodriguez, Taylor, and
  Mota]{commbiol2019}
Yujiang Wang, Joe Necus, Luis~Peraza Rodriguez, Peter~Neal Taylor, and Bruno
  Mota.
\newblock Human cortical folding across regions within individual brains
  follows universal scaling law.
\newblock \emph{Communications Biology}, 2\penalty0 (1):\penalty0 1--8, May
  2019.
\newblock ISSN 2399-3642.
\newblock \doi{10.1038/s42003-019-0421-7}.
\newblock URL \url{https://www.nature.com/articles/s42003-019-0421-7}.

\bibitem[Leiberg et~al.(2021)Leiberg, Papasavvas, and Wang]{leiberg2021}
Karoline Leiberg, Christoforos Papasavvas, and Yujiang Wang.
\newblock Local morphological measures confirm that folding within small
  partitions of the human cortex follows universal scaling law.
\newblock In Marleen de~Bruijne, Philippe~C. Cattin, St{\'e}phane Cotin,
  Nicolas Padoy, Stefanie Speidel, Yefeng Zheng, and Caroline Essert, editors,
  \emph{Medical Image Computing and Computer Assisted Intervention -- MICCAI
  2021}, pages 691--700, Cham, 2021. Springer International Publishing.
\newblock ISBN 978-3-030-87234-2.

\bibitem[Mandelbrot(1983)]{Mandelbrot1983Book}
B.~B. Mandelbrot.
\newblock \emph{The fractal geometry of nature}.
\newblock W. H. Freeman and Comp., New York, 3 edition, 1983.

\bibitem[Elston and Zietsch(2005)]{Elston2005}
G.~N. Elston and Brendan Zietsch.
\newblock Fractal analysis as a tool for studying specialization in neuronal
  structure: the study of the evolution of the primate cerebral cortex and
  human intellect.
\newblock \emph{Adv. Complex Syst.}, 8\penalty0 (2-3):\penalty0 217--227, 2005.
\newblock URL
  \url{http://dblp.uni-trier.de/db/journals/advcs/advcs8.html#ElstonZ05}.

\bibitem[Plank and Benhamou(2008)]{Codling2008}
Michael~J. Plank and Simon Benhamou.
\newblock Random walk models in biology.
\newblock \emph{Journal of the Royal Society, Interface}, 5\penalty0
  (35):\penalty0 21, August 2008.
\newblock ISSN 2198-4026.
\newblock \doi{10.1098/rsif.2008.0014}.
\newblock URL \url{https://doi.org/10.1098/rsif.2008.0014}.

\bibitem[Ionescu et~al.(2009)Ionescu, Oustaloup, Levron, Melchior, Sabatier,
  and {De Keyser}]{IONESCU2009}
Clara Ionescu, Alain Oustaloup, François Levron, Pierre Melchior, Jocelyn
  Sabatier, and Robin {De Keyser}.
\newblock A model of the lungs based on fractal geometrical and structural
  properties.
\newblock \emph{IFAC Proceedings Volumes}, 42\penalty0 (10):\penalty0 994--999,
  2009.
\newblock ISSN 1474-6670.
\newblock \doi{https://doi.org/10.3182/20090706-3-FR-2004.00165}.
\newblock URL
  \url{https://www.sciencedirect.com/science/article/pii/S1474667016387791}.
\newblock 15th IFAC Symposium on System Identification.

\bibitem[Losa(2011)]{Losa2011}
Gabriele~Angelo Losa.
\newblock \emph{Fractals in Biology and Medicine}.
\newblock John Wiley \& Sons, Ltd, 2011.
\newblock ISBN 9783527600908.
\newblock \doi{https://doi.org/10.1002/3527600906.mcb.201100002}.
\newblock URL
  \url{https://onlinelibrary.wiley.com/doi/abs/10.1002/3527600906.mcb.201100002}.

\bibitem[Klonowski(2016)]{Klonowski2016}
Wlodzimierz Klonowski.
\newblock \emph{Fractal Analysis of Electroencephalographic Time Series (EEG
  Signals)}, pages 413--429.
\newblock Springer, 08 2016.
\newblock ISBN 978-1-4939-3993-0.
\newblock \doi{10.1007/978-1-4939-3995-4_25}.

\bibitem[Di~Ieva(2016)]{DiIevabook2016}
Antonio Di~Ieva.
\newblock \emph{The Fractal Geometry of the Brain}.
\newblock Springer Series in Computational Neuroscience. Springer, 2016.
\newblock ISBN 978-1-4939-3995-4.
\newblock \doi{10.1007/978-1-4939-3995-4}.

\bibitem[Reznikov et~al.(2018)Reznikov, Bilton, Lari, Stevens, and
  Kröger]{Reznikov2018}
Natalie Reznikov, Matthew Bilton, Leonardo Lari, Molly~M. Stevens, and Roland
  Kröger.
\newblock Fractal-like hierarchical organization of bone begins at the
  nanoscale.
\newblock \emph{Science}, 360\penalty0 (6388):\penalty0 eaao2189, 2018.
\newblock \doi{10.1126/science.aao2189}.
\newblock URL \url{https://www.science.org/doi/abs/10.1126/science.aao2189}.

\bibitem[Free et~al.(1996)Free, Sisodiya, Cook, Fish, and Shorvon]{Free1996}
S.~L. Free, S.~M. Sisodiya, M.~J. Cook, D.~R. Fish, and S.~D. Shorvon.
\newblock {Three-Dimensional Fractal Analysis of the White Matter Surface from
  Magnetic Resonance Images of the Human Brain}.
\newblock \emph{Cerebral Cortex}, 6\penalty0 (6):\penalty0 830--836, 11 1996.
\newblock ISSN 1047-3211.
\newblock \doi{10.1093/cercor/6.6.830}.
\newblock URL \url{https://doi.org/10.1093/cercor/6.6.830}.

\bibitem[Kiselev et~al.(2003)Kiselev, Hahn, and Auer]{KISELEV2003}
Valerij~G. Kiselev, Klaus~R. Hahn, and Dorothee~P. Auer.
\newblock Is the brain cortex a fractal?
\newblock \emph{NeuroImage}, 20\penalty0 (3):\penalty0 1765--1774, 2003.
\newblock ISSN 1053-8119.
\newblock \doi{https://doi.org/10.1016/S1053-8119(03)00380-X}.
\newblock URL
  \url{https://www.sciencedirect.com/science/article/pii/S105381190300380X}.

\bibitem[Marzi et~al.(2021)Marzi, Giannelli, Tessa, Mascalchi, and
  Diciotti]{Marzi2021}
Chiara Marzi, Marco Giannelli, Carlo Tessa, Mario Mascalchi, and Stefano
  Diciotti.
\newblock Fractal analysis of mri data at 7 t: How much complex is the cerebral
  cortex?
\newblock \emph{IEEE Access}, 9:\penalty0 69226--69234, 2021.
\newblock \doi{10.1109/ACCESS.2021.3077370}.

\bibitem[Jao et~al.(2021)Jao, Lau, Lien, Tsai, Chu, Hsiao, Yeh, and
  Wu]{Jao2021}
Chi-Wen Jao, Chi~Ieong Lau, Li-Ming Lien, Yuh-Feng Tsai, Kuang-En Chu, Chen-Yu
  Hsiao, Jiann-Horng Yeh, and Yu-Te Wu.
\newblock Using fractal dimension analysis with the desikan–killiany atlas to
  assess the effects of normal aging on subregional cortex alterations in
  adulthood.
\newblock \emph{Brain Sciences}, 11\penalty0 (1), 2021.
\newblock ISSN 2076-3425.
\newblock \doi{10.3390/brainsci11010107}.
\newblock URL \url{https://www.mdpi.com/2076-3425/11/1/107}.

\bibitem[Meregalli et~al.(2022)Meregalli, Alberti, Madan, Meneguzzo, Miola,
  Trevisan, Sambataro, Favaro, and Collantoni]{Meregalli2022}
Valentina Meregalli, Francesco Alberti, Christopher~R. Madan, Paolo Meneguzzo,
  Alessandro Miola, Nicol{\`{o}} Trevisan, Fabio Sambataro, Angela Favaro, and
  Enrico Collantoni.
\newblock Cortical complexity estimation using fractal dimension: A systematic
  review of the literature on clinical and nonclinical samples.
\newblock \emph{European Journal of Neuroscience}, 55\penalty0 (6):\penalty0
  1547--1583, 2022.
\newblock \doi{10.1111/ejn.15631}.

\bibitem[King et~al.(2010)King, Brown, Hwang, Jeon, and George]{King2010}
Richard~D. King, Brandon Brown, Michael Hwang, Tina Jeon, and Anuh~T. George.
\newblock Fractal dimension analysis of the cortical ribbon in mild
  {Alzheimer’s} disease.
\newblock \emph{NeuroImage}, 53\penalty0 (2):\penalty0 471--479, 2010.
\newblock \doi{10.1016/j.neuroimage.2010.06.050}.

\bibitem[Madan and Kensinger(2016)]{madan_cortical_2016}
Christopher~R. Madan and Elizabeth~A. Kensinger.
\newblock Cortical complexity as a measure of age-related brain atrophy.
\newblock \emph{NeuroImage}, 134\penalty0 (Supplement C):\penalty0 617--629,
  July 2016.
\newblock ISSN 1053-8119.
\newblock \doi{10.1016/j.neuroimage.2016.04.029}.
\newblock URL
  \url{http://www.sciencedirect.com/science/article/pii/S1053811916300519}.

\bibitem[Marzi et~al.(2020)Marzi, Giannelli, Tessa, Mascalchi, and
  Diciotti]{Marzi2020}
C.~Marzi, M.~Giannelli, C.~Tessa, M.~Mascalchi, and S.~Diciotti.
\newblock Toward a more reliable characterization of fractal properties of the
  cerebral cortex of healthy subjects during the lifespan.
\newblock \emph{Sci Rep}, 10\penalty0 (16957), 2020.
\newblock \doi{10.1038/s41598-020-73961-w}.
\newblock URL \url{https://www.nature.com/articles/s41598-020-73961-w#content}.

\bibitem[Kochunov et~al.(2007)Kochunov, Thompson, Coyle, Lancaster, Kochunov,
  Royall, Mangin, Rivière, and Fox]{Kochunov2007}
Peter Kochunov, Paul~M. Thompson, Thomas~R. Coyle, Jack~L. Lancaster, Valeria
  Kochunov, Donal Royall, Jean‐Frans\c{c}ois Mangin, Denis Rivière, and
  Peter~T. Fox.
\newblock Relationship among neuroimaging indices of cerebral health during
  normal aging.
\newblock \emph{Human Brain Mapping}, 29\penalty0 (1):\penalty0 36--45, 2007.
\newblock \doi{10.1002/hbm.20369}.

\bibitem[Madan(2019)]{Madan2019}
Christopher~R. Madan.
\newblock Robust estimation of sulcal morphology.
\newblock \emph{Brain Informatics}, 6\penalty0 (1):\penalty0 5, June 2019.
\newblock ISSN 2198-4026.
\newblock \doi{10.1186/s40708-019-0098-1}.
\newblock URL \url{https://doi.org/10.1186/s40708-019-0098-1}.

\bibitem[Wang et~al.(2021)Wang, Leiberg, Ludwig, Little, Necus, Winston, Vos,
  Tisi, Duncan, Taylor, and Mota]{wang2021}
Yujiang Wang, Karoline Leiberg, Tobias Ludwig, Bethany Little, Joe~H Necus,
  Gavin Winston, Sjoerd~B Vos, Jane~de Tisi, John~S Duncan, Peter~N Taylor, and
  Bruno Mota.
\newblock Independent components of human brain morphology.
\newblock \emph{NeuroImage}, 226:\penalty0 117546, February 2021.
\newblock ISSN 1053-8119.
\newblock \doi{10.1016/j.neuroimage.2020.117546}.
\newblock URL
  \url{https://www.sciencedirect.com/science/article/pii/S1053811920310314}.

\bibitem[Van~Essen et~al.(2012)Van~Essen, Ugurbil, Auerbach, Barch, Behrens,
  Bucholz, Chang, Chen, Corbetta, Curtiss, Della~Penna, Feinberg, Glasser,
  Harel, Heath, Larson-Prior, Marcus, Michalareas, Moeller, Oostenveld,
  Petersen, Prior, Schlaggar, Smith, Snyder, Xu, and Yacoub]{van_essen2012}
D.~C. Van~Essen, K.~Ugurbil, E.~Auerbach, D.~Barch, T.~E.~J. Behrens,
  R.~Bucholz, A.~Chang, L.~Chen, M.~Corbetta, S.~W. Curtiss, S.~Della~Penna,
  D.~Feinberg, M.~F. Glasser, N.~Harel, A.~C. Heath, L.~Larson-Prior,
  D.~Marcus, G.~Michalareas, S.~Moeller, R.~Oostenveld, S.~E. Petersen,
  F.~Prior, B.~L. Schlaggar, S.~M. Smith, A.~Z. Snyder, J.~Xu, and E.~Yacoub.
\newblock The {Human} {Connectome} {Project}: {A} data acquisition perspective.
\newblock \emph{NeuroImage}, 62\penalty0 (4):\penalty0 2222--2231, October
  2012.
\newblock ISSN 1053-8119.
\newblock \doi{10.1016/j.neuroimage.2012.02.018}.
\newblock URL
  \url{http://www.sciencedirect.com/science/article/pii/S1053811912001954}.

\bibitem[Shafto et~al.(2014)Shafto, Tyler, Dixon, Taylor, Rowe, Cusack, Calder,
  Marslen-Wilson, Duncan, Dalgleish, et~al.]{shafto2014cambridge}
Meredith~A Shafto, Lorraine~K Tyler, Marie Dixon, Jason~R Taylor, James~B Rowe,
  Rhodri Cusack, Andrew~J Calder, William~D Marslen-Wilson, John Duncan, Tim
  Dalgleish, et~al.
\newblock The cambridge centre for ageing and neuroscience (cam-can) study
  protocol: a cross-sectional, lifespan, multidisciplinary examination of
  healthy cognitive ageing.
\newblock \emph{BMC neurology}, 14\penalty0 (1):\penalty0 204, 2014.

\bibitem[Taylor et~al.(2017)Taylor, Williams, Cusack, Auer, Shafto, Dixon,
  Tyler, Henson, et~al.]{taylor2017cambridge}
Jason~R Taylor, Nitin Williams, Rhodri Cusack, Tibor Auer, Meredith~A Shafto,
  Marie Dixon, Lorraine~K Tyler, Richard~N Henson, et~al.
\newblock The cambridge centre for ageing and neuroscience (cam-can) data
  repository: structural and functional mri, meg, and cognitive data from a
  cross-sectional adult lifespan sample.
\newblock \emph{Neuroimage}, 144:\penalty0 262--269, 2017.

\bibitem[Nooner et~al.(2012)Nooner, Colcombe, Tobe, Mennes, Benedict, Moreno,
  Panek, Brown, Zavitz, Li, Sikka, Gutman, Bangaru, Schlachter, Kamiel, Anwar,
  Hinz, Kaplan, Rachlin, Adelsberg, Cheung, Khanuja, Yan, Craddock, Calhoun,
  Courtney, King, Wood, Cox, Kelly, DiMartino, Petkova, Reiss, Duan, Thompsen,
  Biswal, Coffey, Hoptman, Javitt, Pomara, Sidtis, Koplewicz, Castellanos,
  Leventhal, and Milham]{Nooner2012}
Kate Nooner, Stanley Colcombe, Russell Tobe, Maarten Mennes, Melissa Benedict,
  Alexis Moreno, Laura Panek, Shaquanna Brown, Stephen Zavitz, Qingyang Li,
  Sharad Sikka, David Gutman, Saroja Bangaru, Rochelle~Tziona Schlachter,
  Stephanie Kamiel, Ayesha Anwar, Caitlin Hinz, Michelle Kaplan, Anna Rachlin,
  Samantha Adelsberg, Brian Cheung, Ranjit Khanuja, Chaogan Yan, Cameron
  Craddock, Vincent Calhoun, William Courtney, Margaret King, Dylan Wood,
  Christine Cox, Clare Kelly, Adriana DiMartino, Eva Petkova, Philip Reiss,
  Nancy Duan, Dawn Thompsen, Bharat Biswal, Barbara Coffey, Matthew Hoptman,
  Daniel Javitt, Nunzio Pomara, John Sidtis, Harold Koplewicz, Francisco
  Castellanos, Bennett Leventhal, and Michael Milham.
\newblock The nki-rockland sample: A model for accelerating the pace of
  discovery science in psychiatry.
\newblock \emph{Frontiers in Neuroscience}, 6, 2012.
\newblock ISSN 1662-453X.
\newblock \doi{10.3389/fnins.2012.00152}.
\newblock URL
  \url{https://www.frontiersin.org/article/10.3389/fnins.2012.00152}.

\bibitem[Liu et~al.(2018)Liu, Ye, Yen, Newman, Glen, Leopold, and
  Silva]{LiuEtal2018}
Cirong Liu, Frank~Q. Ye, Cecil Chern-Chyi Yen, John~D. Newman, Daniel Glen,
  David~A. Leopold, and Afonso~C. Silva.
\newblock A digital {3D} atlas of the marmoset brain based on multi-modal
  {MRI}.
\newblock \emph{{NeuroImage}}, 169:\penalty0 106--116, 2018.
\newblock \doi{10.1016/j.neuroimage.2017.12.004}.

\bibitem[Liu et~al.(2020)Liu, Ye, Newman, Szczupak, Tian, Yen, Majka, Glen,
  Rosa, Leopold, and Silva]{LiuEtal2020}
Cirong Liu, Frank~Q. Ye, John~D. Newman, Diego Szczupak, Xiaoguang Tian, Cecil
  Chern-Chyi Yen, Piotr Majka, Daniel Glen, Marcello G.~P. Rosa, David~A.
  Leopold, and Afonso~C. Silva.
\newblock A resource for the detailed 3d mapping of white matter pathways in
  the marmoset brain.
\newblock \emph{Nature Neuroscience}, 23\penalty0 (2):\penalty0 271--280, 2020.
\newblock \doi{10.1038/s41593-019-0575-0}.

\bibitem[Madan(2015)]{Madan2015}
Christopher~R. Madan.
\newblock Creating {3D} visualizations of {MRI} data: A brief guide.
\newblock \emph{F1000Research}, 4:\penalty0 466, 2015.
\newblock \doi{10.12688/f1000research.6838.1}.

\bibitem[Madan(2016)]{Madan2016}
Christopher~R Madan.
\newblock Improved understanding of brain morphology through 3d printing: A
  brief guide.
\newblock \emph{Research Ideas and Outcomes}, 2:\penalty0 e10398, 2016.
\newblock \doi{10.3897/rio.2.e10398}.

\bibitem[Ardesch et~al.(2022)Ardesch, Scholtens, {de Lange}, Roumazeilles,
  Khrapitchev, Preuss, Rilling, Mars, and {van den Heuvel}]{ardesch2022}
Dirk~Jan Ardesch, Lianne~H Scholtens, Siemon~C {de Lange}, Lea Roumazeilles,
  Alexandre~A Khrapitchev, Todd~M Preuss, James~K Rilling, Rogier~B Mars, and
  Martijn~P {van den Heuvel}.
\newblock Scaling {{Principles}} of {{White Matter Connectivity}} in the
  {{Human}} and {{Nonhuman Primate Brain}}.
\newblock \emph{Cerebral Cortex}, 32\penalty0 (13):\penalty0 2831--2842, July
  2022.
\newblock ISSN 1047-3211.
\newblock \doi{10.1093/cercor/bhab384}.

\bibitem[Bryant et~al.(2021)Bryant, Ardesch, Roumazeilles, Scholtens,
  Khrapitchev, Tendler, Wu, Miller, Sallet, {van den Heuvel}, and
  Mars]{bryant2021}
Katherine~L. Bryant, Dirk~Jan Ardesch, Lea Roumazeilles, Lianne~H. Scholtens,
  Alexandre~A. Khrapitchev, Benjamin~C. Tendler, Wenchuan Wu, Karla~L. Miller,
  Jerome Sallet, Martijn~P. {van den Heuvel}, and Rogier~B. Mars.
\newblock Diffusion {{MRI}} data, sulcal anatomy, and tractography for eight
  species from the {{Primate Brain Bank}}.
\newblock \emph{Brain Structure and Function}, 226\penalty0 (8):\penalty0
  2497--2509, November 2021.
\newblock ISSN 1863-2661.
\newblock \doi{10.1007/s00429-021-02268-x}.

\bibitem[Jin et~al.(2018)Jin, Zhang, Shaw, Sachdev, and Cherbuin]{Jin2018}
Kaide Jin, Tianqi Zhang, Marnie Shaw, Perminder Sachdev, and Nicolas Cherbuin.
\newblock Relationship {Between} {Sulcal} {Characteristics} and {Brain}
  {Aging}.
\newblock \emph{Frontiers in Aging Neuroscience}, 10:\penalty0 339, November
  2018.
\newblock ISSN 1663-4365.
\newblock \doi{10.3389/fnagi.2018.00339}.
\newblock URL \url{https://www.ncbi.nlm.nih.gov/pmc/articles/PMC6240579/}.

\bibitem[Leiberg et~al.(2024)Leiberg, Blattner, Little, Mello, de~Moraes,
  Rummel, Taylor, Mota, and Wang]{leiberg2024multiscale}
Karoline Leiberg, Timo Blattner, Bethany Little, Victor B.~B. Mello, Fernanda
  H.~P. de~Moraes, Christian Rummel, Peter~N. Taylor, Bruno Mota, and Yujiang
  Wang.
\newblock Multiscale cortical morphometry reveals pronounced regional and
  scale-dependent variations across the lifespan, 2024.

\bibitem[Hofman(1985)]{Hofman1985}
Michel~A. Hofman.
\newblock Size and shape of the cerebral cortex in mammals.
\newblock \emph{Brain, Behavior and Evolution}, 27\penalty0 (1):\penalty0
  28--40, 1985.
\newblock \doi{10.1159/000118718}.

\bibitem[Hofman(1991)]{Hofman1991}
M~A Hofman.
\newblock The fractal geometry of convoluted brains.
\newblock \emph{Journal f\"ur Hirnforschung}, 32:\penalty0 103--111, 1991.

\bibitem[Xu et~al.(2010)Xu, Knutsen, Dikranian, Kroenke, Bayly, and
  Taber]{xu_axons_2010}
Gang Xu, Andrew~K. Knutsen, Krikor Dikranian, Christopher~D. Kroenke, Philip~V.
  Bayly, and Larry~A. Taber.
\newblock Axons {Pull} on the {Brain}, {But} {Tension} {Does} {Not} {Drive}
  {Cortical} {Folding}.
\newblock \emph{Journal of biomechanical engineering}, 132\penalty0
  (7):\penalty0 071013, July 2010.
\newblock ISSN 0148-0731.
\newblock \doi{10.1115/1.4001683}.
\newblock URL \url{http://www.ncbi.nlm.nih.gov/pmc/articles/PMC3170872/}.

\bibitem[Barenblatt(1996)]{barenblatt1996scaling}
G.I. Barenblatt.
\newblock \emph{Scaling, Self-similarity, and Intermediate Asymptotics:
  Dimensional Analysis and Intermediate Asymptotics}.
\newblock Cambridge Texts in Applied Mathematics. Cambridge University Press,
  1996.
\newblock ISBN 9780521435222.
\newblock URL \url{https://books.google.com.br/books?id=r-Az53e-MTYC}.

\bibitem[West et~al.(1997)West, Brown, and Enquist]{West1997}
Geoffrey~B. West, James~H. Brown, and Brian~J. Enquist.
\newblock {A General Model for the Origin of Allometric Scaling Laws in
  Biology}.
\newblock \emph{Science}, 276\penalty0 (5309):\penalty0 122, March 1997.
\newblock \doi{https://doi.org/10.1126/science.276.5309.122}.
\newblock URL \url{https://www.science.org/doi/10.1126/science.276.5309.122}.

\bibitem[Gagler et~al.(2022)Gagler, Karas, Kempes, Malloy, Mierzejewski,
  Goldman, Kim, and Walker]{gagler_scaling_2022}
Dylan~C. Gagler, Bradley Karas, Christopher~P. Kempes, John Malloy, Veronica
  Mierzejewski, Aaron~D. Goldman, Hyunju Kim, and Sara~I. Walker.
\newblock Scaling laws in enzyme function reveal a new kind of biochemical
  universality.
\newblock \emph{Proceedings of the National Academy of Sciences}, 119\penalty0
  (9):\penalty0 e2106655119, March 2022.
\newblock \doi{10.1073/pnas.2106655119}.
\newblock URL \url{https://www.pnas.org/doi/10.1073/pnas.2106655119}.
\newblock Publisher: Proceedings of the National Academy of Sciences.

\bibitem[Johnston et~al.(2022)Johnston, Dingle, Greenbury, Camargo, Doye,
  Ahnert, and Louis]{Johnston2022}
Iain~G. Johnston, Kamaludin Dingle, Sam~F. Greenbury, Chico~Q. Camargo,
  Jonathan P.~K. Doye, Sebastian~E. Ahnert, and Ard~A. Louis.
\newblock Symmetry and simplicity spontaneously emerge from the algorithmic
  nature of evolution.
\newblock \emph{Proceedings of the National Academy of Sciences}, 119\penalty0
  (11):\penalty0 e2113883119, 2022.
\newblock \doi{10.1073/pnas.2113883119}.
\newblock URL \url{https://www.pnas.org/doi/abs/10.1073/pnas.2113883119}.

\bibitem[Quezada et~al.(2020)Quezada, van~de Looij, Hale, Rana, Sizonenko,
  Gilchrist, Castillo-Melendez, Tolcos, and Walker]{Quezada2020}
Sebastian Quezada, Yohan van~de Looij, Nadia Hale, Shreya Rana, Stéphane~V
  Sizonenko, Courtney Gilchrist, Margie Castillo-Melendez, Mary Tolcos, and
  David~W Walker.
\newblock {Genetic and microstructural differences in the cortical plate of
  gyri and sulci during gyrification in fetal sheep}.
\newblock \emph{Cerebral Cortex}, 30\penalty0 (12):\penalty0 6169--6190, 07
  2020.
\newblock ISSN 1047-3211.
\newblock \doi{10.1093/cercor/bhaa171}.
\newblock URL \url{https://doi.org/10.1093/cercor/bhaa171}.

\bibitem[Molnár et~al.(2014)Molnár, Kaas, de~Carlos, Hevner, Lein, and
  Němec]{molnar2014}
Zoltán Molnár, Jon~H. Kaas, Juan~A. de~Carlos, Robert~F. Hevner, Ed~Lein, and
  Pavel Němec.
\newblock Evolution and development of the mammalian cerebral cortex.
\newblock \emph{Brain, Behavior and Evolution}, 83\penalty0 (2):\penalty0
  126–139, 2014.
\newblock \doi{10.1159/000357753}.

\bibitem[Kaas(2012)]{Kaas2012}
Jon~H. Kaas.
\newblock Evolution of columns, modules, and domains in the neocortex of
  primates.
\newblock \emph{Proceedings of the National Academy of Sciences}, 109\penalty0
  (supplement\_1):\penalty0 10655--10660, 2012.
\newblock \doi{10.1073/pnas.1201892109}.
\newblock URL \url{https://www.pnas.org/doi/abs/10.1073/pnas.1201892109}.

\bibitem[Zilles et~al.(2013)Zilles, Palomero-Gallagher, and Amunts]{ZILLES2013}
Karl Zilles, Nicola Palomero-Gallagher, and Katrin Amunts.
\newblock Development of cortical folding during evolution and ontogeny.
\newblock \emph{Trends in Neurosciences}, 36\penalty0 (5):\penalty0 275--284,
  2013.
\newblock ISSN 0166-2236.
\newblock \doi{https://doi.org/10.1016/j.tins.2013.01.006}.
\newblock URL
  \url{https://www.sciencedirect.com/science/article/pii/S0166223613000180}.

\bibitem[Garcia et~al.(2018)Garcia, Robinson, Alexopoulos, Dierker, Glasser,
  Coalson, Ortinau, Rueckert, Taber, Essen, Rogers, Smyser, and
  Bayly]{Garcia2018}
Kara~E. Garcia, Emma~C. Robinson, Dimitrios Alexopoulos, Donna~L. Dierker,
  Matthew~F. Glasser, Timothy~S. Coalson, Cynthia~M. Ortinau, Daniel Rueckert,
  Larry~A. Taber, David C.~Van Essen, Cynthia~E. Rogers, Christopher~D. Smyser,
  and Philip~V. Bayly.
\newblock Dynamic patterns of cortical expansion during folding of the preterm
  human brain.
\newblock \emph{Proceedings of the National Academy of Sciences}, 115\penalty0
  (12):\penalty0 3156--3161, 2018.
\newblock \doi{10.1073/pnas.1715451115}.
\newblock URL \url{https://www.pnas.org/doi/abs/10.1073/pnas.1715451115}.

\bibitem[Pizzagalli et~al.(2020)Pizzagalli, Auzias, Yang, Mathias, Faskowitz,
  Boyd, Amini, Rivi{\`e}re, McMahon, de~Zubicaray, Martin, Mangin, Glahn,
  Blangero, Wright, Thompson, Kochunov, and Jahanshad]{Pizzagalli2020}
Fabrizio Pizzagalli, Guillaume Auzias, Qifan Yang, Samuel~R. Mathias, Joshua
  Faskowitz, Joshua Boyd, Armand Amini, Denis Rivi{\`e}re, Katie~L. McMahon,
  Greig~I. de~Zubicaray, Nicholas~G. Martin, Jean-Fran{\c c}ois Mangin,
  David~C. Glahn, John Blangero, Margaret~J. Wright, Paul~M. Thompson, Peter
  Kochunov, and Neda Jahanshad.
\newblock The reliability and heritability of cortical folds and their genetic
  correlations across hemispheres.
\newblock \emph{Communications Biology}, 2020.
\newblock \doi{10.1038/s42003-020-01163-1}.
\newblock URL \url{https://doi.org/10.1038/s42003-020-01163-1}.

\bibitem[Heuer et~al.(2019)Heuer, Gulban, Bazin, Osoianu, Valabregue, Santin,
  Herbin, and Toro]{HEUER2019}
Katja Heuer, Omer~Faruk Gulban, Pierre-Louis Bazin, Anastasia Osoianu, Romain
  Valabregue, Mathieu Santin, Marc Herbin, and Roberto Toro.
\newblock Evolution of neocortical folding: A phylogenetic comparative analysis
  of mri from 34 primate species.
\newblock \emph{Cortex}, 118:\penalty0 275--291, 2019.
\newblock ISSN 0010-9452.
\newblock \doi{https://doi.org/10.1016/j.cortex.2019.04.011}.
\newblock URL
  \url{https://www.sciencedirect.com/science/article/pii/S0010945219301704}.
\newblock The Evolution of the Mind and the Brain.

\bibitem[Valk et~al.(2020)Valk, Xu, Margulies, Masouleh, Paquola, Goulas,
  Kochunov, Smallwood, Yeo, Bernhardt, and Eickhoff]{Valk2020}
Sofie~L. Valk, Ting Xu, Daniel~S. Margulies, Shahrzad~Kharabian Masouleh, Casey
  Paquola, Alexandros Goulas, Peter Kochunov, Jonathan Smallwood, B.~T.~Thomas
  Yeo, Boris~C. Bernhardt, and Simon~B. Eickhoff.
\newblock Shaping brain structure: {Genetic} and phylogenetic axes of
  macroscale organization of cortical thickness.
\newblock \emph{Science Advances}, 6\penalty0 (39):\penalty0 eabb3417,
  September 2020.
\newblock ISSN 2375-2548.
\newblock \doi{10.1126/sciadv.abb3417}.
\newblock URL \url{https://advances.sciencemag.org/content/6/39/eabb3417}.
\newblock Publisher: American Association for the Advancement of Science
  Section: Research Article.

\bibitem[Mars et~al.(2014)Mars, Neubert, Verhagen, Sallet, Miller, Dunbar, and
  Barton]{Mars2014}
Rogier~B. Mars, Franz-Xaver Neubert, Lennart Verhagen, Jerome Sallet, Karla~L.
  Miller, Robin~I. Dunbar, and Robert~A. Barton.
\newblock Primate comparative neuroscience using magnetic resonance imaging:
  Promises and challenges.
\newblock \emph{Frontiers in Neuroscience}, 8, 2014.
\newblock \doi{10.3389/fnins.2014.00298}.

\bibitem[Croxson et~al.(2017)Croxson, Forkel, Cerliani, and Thiebaut~de
  Schotten]{Croxson2017}
Paula~L Croxson, Stephanie~J Forkel, Leonardo Cerliani, and Michel Thiebaut~de
  Schotten.
\newblock {Structural Variability Across the Primate Brain: A Cross-Species
  Comparison}.
\newblock \emph{Cerebral Cortex}, 28\penalty0 (11):\penalty0 3829--3841, 10
  2017.
\newblock ISSN 1047-3211.
\newblock \doi{10.1093/cercor/bhx244}.
\newblock URL \url{https://doi.org/10.1093/cercor/bhx244}.

\bibitem[Schaer et~al.(2008)Schaer, Cuadra, Tamarit, Lazeyras, Eliez, and
  Thiran]{schaer_surface-based_2008}
M.~Schaer, M.~B. Cuadra, L.~Tamarit, F.~Lazeyras, S.~Eliez, and J.~P. Thiran.
\newblock A {Surface}-{Based} {Approach} to {Quantify} {Local} {Cortical}
  {Gyrification}.
\newblock \emph{IEEE Transactions on Medical Imaging}, 27\penalty0
  (2):\penalty0 161--170, February 2008.
\newblock ISSN 0278-0062.
\newblock \doi{10.1109/TMI.2007.903576}.

\bibitem[Yu et~al.(2021)Yu, Brakensiek, Schumacher, and Crane]{Yu2021}
Chris Yu, Caleb Brakensiek, Henrik Schumacher, and Keenan Crane.
\newblock Repulsive surfaces.
\newblock \emph{ACM Trans. Graph.}, 40\penalty0 (6), dec 2021.
\newblock ISSN 0730-0301.
\newblock \doi{10.1145/3478513.3480521}.
\newblock URL \url{https://doi.org/10.1145/3478513.3480521}.

\bibitem[Raznahan et~al.(2011)Raznahan, Shaw, Lalonde, Stockman, Wallace,
  Greenstein, Clasen, Gogtay, and Giedd]{Raznahan2011}
Armin Raznahan, Phillip Shaw, Francois Lalonde, Mike Stockman, Gregory~L.
  Wallace, Dede Greenstein, Liv Clasen, Nitin Gogtay, and Jay~N. Giedd.
\newblock How {Does} {Your} {Cortex} {Grow}?
\newblock \emph{The Journal of Neuroscience}, 31\penalty0 (19):\penalty0
  7174--7177, May 2011.
\newblock ISSN 0270-6474.
\newblock \doi{10.1523/JNEUROSCI.0054-11.2011}.
\newblock URL \url{https://www.ncbi.nlm.nih.gov/pmc/articles/PMC3157294/}.

\end{thebibliography}

\section*{Acknowledgements}
We thank members of the Computational Neurology, Neuroscience \& Psychiatry Lab (www.cnnp-lab.com) for discussions on the analysis and manuscript, and Dirk Jan Ardesch and Martijn van den Heuvel for helpful discussions and NHP brain surface data. P.N.T. and Y.W. are both supported by UKRI Future Leaders Fellowships (MR/T04294X/1, MR/V026569/1); Y.W. and K.L are further supported by the EPSRC (EP/Y016009/1, EP/L015358/1). B. Mota is supported by Fundação Serrapilheira Institute (grant Serra-1709-16981) and CNPq (PQ 2017 312837/2017-8).

\newpage
\renewcommand{\thefigure}{S\arabic{figure}}
\setcounter{figure}{0}
\counterwithin{figure}{section}
\counterwithin{table}{section}
\renewcommand\thesection{S\arabic{section}}
\setcounter{section}{0}

\section*{Supplementary}

\section{Coarse-graining algorithm\label{coarsegrainingalgo}}

The core coarse-graining algorithm underpinning our analyses takes pial and white matter surfaces from the cortical surface reconstruction as inputs and at any specified spatial scale $\lambda$ ``voxelises'' the gray matter ribbon at the specified resolution. To achieve this, we set up a 3D voxel grid, where each grid voxel is of size $\lambda \times \lambda \times \lambda$.

In detail, we assign all voxels in the grid with at least four corners inside the original pial surface to the pial voxelization. This process allows the exposed surface to remain approximately constant with increasing voxel sizes. A constant exposed surface is desirable, as we only want to gradually ``melt'' and fuse the gyri, but not grow the bounding/exposed surface as well. We want the extrinsic area to remain approximately constant as we decrease the instrinsic area via coarse-graining; it is like generating iterates of a Koch curve in reverse, from more to less detailed, by increasing the length of smallest line segment. 

We then assign voxels with all eight corners inside the original white matter surface to the white matter voxelization. This is to ensure integrity of the white matter, as otherwise white matter voxels in gyri may become detached from the core white matter, and thus artificially increase white matter surface area. Indeed the main results of the paper are not very sensitive to this decision using all eight corners, vs. e.g. only four corners, as we do not directly use white matter surface area for the scaling law measurements. However, we still maintained this choice in case future work wants to make use of the white matter voxelisations or derivative measures.

Finally, subtracting the white matter voxelisation from the pial voxelisation, we obtain the grey matter voxelisation (Fig.~\ref{fig:meltingprocess}B bottom row). The pial surface voxelisation is designed such that neighbouring gyri can fuse if their separation is smaller than the scale of interest, whilst not growing the cortex outwards. The white matter surface voxelisation allows the walls of each gyrus to thicken and fuse inwards. Visually, this process looks as if the cortex is ``thickening'' inwards and smoothing on the outside.

From the voxelised cortex at each scale $\lambda$, we can then obtain coarse-grained pial and white matter surfaces, and extract estimates of global morphometric measures such as the average cortical thickness $T(\lambda)$, the total cortical surface area $A_t(\lambda)$, and the exposed surface area $A_e(\lambda)$. For each $\lambda$, we derived an outer isosurface equal to 0.5 for the pial voxelisation (voxel values are 1 within the voxelised pial surface and 0 otherwise). This isosurface is then defined as the coarse-grained pial surface at this scale. The surface area of this isosurface is used as $A_t(\lambda)$. The exposed surface area $A_e(\lambda)$ is subsequently derived from the convex hull of the pial isosurface. Finally, the cortical thickness $T(\lambda)$ is estimated as $\frac{V_G(\lambda)}{A_t(\lambda)}$, where $V_G(\lambda)$, which is the estimated grey matter volume, derived from the number of grey matter voxels, is multiplied by the voxel volume. These global morphometric measures are summary statistics of the brain at each particular scale, capturing information about both their intrinsic geometry ($A_t(\lambda), T(\lambda)$) and extrinsic geometry ($A_e(\lambda)$). In this manner, for each cortex, we obtain not just one set of summary morphometric measures, but rather an set measures as functions of $\lambda$.
A detailed walk-through of the coarse-graining algorithm and estimation of morphometric measures is provided on Github: \url{https://github.com/cnnp-lab/CorticalFoldingAnalysisTools/blob/master/Scales/fastEstimateScale.m}.

Note that although this process is inspired by the box-counting algorithm, it is different from box-counting and related convolution-based algorithms: we do not simply apply successive convolutions with an increasing kernel size (or equivalent), which effectively would achieve a uniform and spatially isometric ``smearing'' of the original cortical ribbon to a given scale, rather than a targeted erasure of \textit{surface} details smaller than said scale. As a result, this method yields well-defined and well-behaved white and gray matter surfaces. Thus, one can apply all the usual analytical tools to these realisations that are applicable to actual cortices. Note that our approach is more comparable in its principles to the calculation of the outer smoothed pial surface in FreeSurfer \citep{schaer_surface-based_2008}, which utilises a dilation \textit{and} erosion convolution to effectively erase details below a certain scale. Of course, our proposed procedure is not the only conceivable way to erase morphological details below a given scale; and we are actively working on related algorithms that are also computationally cheaper. Nevertheless, the current version requires no fine-tuning, is computationally feasible and conceptually simple, thus making it a natural choice for introducing the methodology and approach. 

Given how broadly it has been verified, we expect the observed universality in cortical self-similar scaling to be robust to the details of the coarse-graining algorithm. It would be very informative to test this proposition in future. For example, an alternative method inspired by \citep{Yu2021} could be implemented, eschewing voxelization and dealing only with the flow of nested surfaces with self- and mutual- avoidance explicitly implemented. Going in the opposite direction, more detailed models for the mechanisms of cortical folding (see e.g.  \citep{Raznahan2011}) can be regarded as a type of reverse melting, and could perhaps be implemented and tested in a similar fashion as fine-graining procedures.

\section{Influence of original image resolution \label{scan_res}}
\begin{figure}[h!]
\begin{adjustwidth}{-1cm}{0cm}
    \centering
\includegraphics{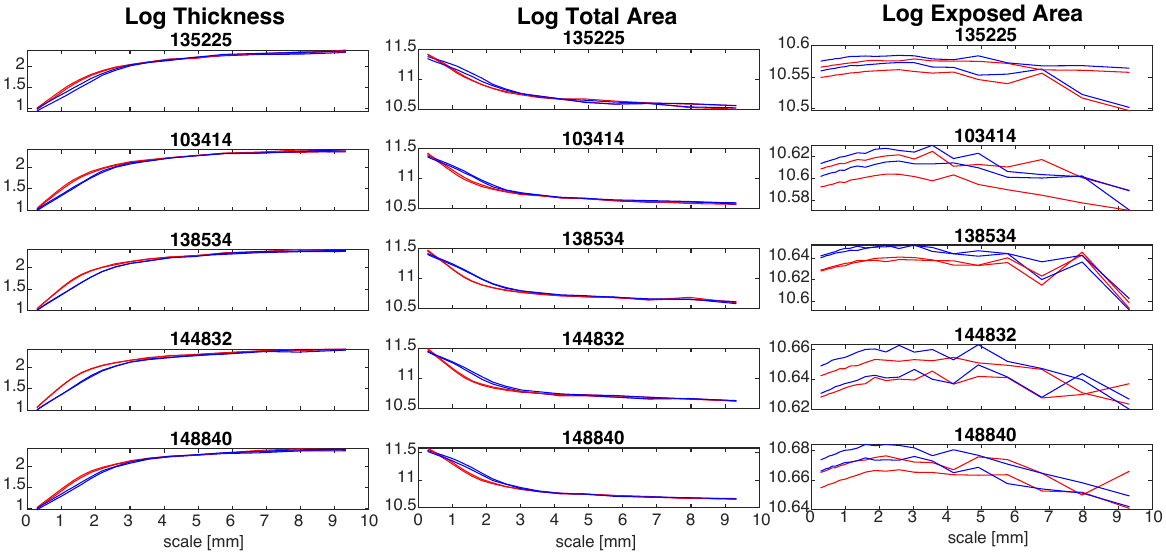}
\end{adjustwidth}
\caption{\textbf{Effect of using different original image resolution in five example HCP subjects.} In all plots, red lines indicate an original image resolution of 0.7mm isotropic, whereas blue line indicate an original image resolution of 1mm isotropic. Both hemispheres are shown for each subject.} \label{sfig:scan_res}

\end{figure}

In this supplementary section, we want to demonstrate the relatively weak effect of the original image resolution on our analysis outputs. To this end, we used five example HCP subjects, who were scanned at 0.7mm isotropic image resolution, and downsampled their images to 1mm isotropic images. We then proceeded with our analysis using two freesurfer outputs. (1) the HCP freesurfer pipeline output optimised for the 0.7mm resolution, and (2) a standard freesurfer pipeline output on the 1mm downsampled images. We proceeded with these two sets of surfaces in our analysis and show the resulting morphology measures in Fig.~\ref{sfig:scan_res}. 

We observe a relatively weak difference between these two sets of inputs/surfaces. Both sets largely follow the same trajectory across scales. Especially in the exposed area, the within and between subject differences are noticeably larger than between image resolutions. In total area and thickness, some small but systematic differences are seen in all subjects between the scales of 1-2mm. These are most likely differences in the cortical morphology reconstruction in the freesurfer surfaces using the two different resolution images as input. However, in none of these resulting morphology measures do we see an artifact specifically at 0.7 or 1mm. Any differences between image resolutions only result in subtle changes in the freesurfer meshes, which are smooth. \textit{Our analysis method therefore has no direct dependency on the image resolution}, as our inputs are these smooth freesurfer meshes.

\section{Scaling properties by species \label{suppl_species_detail}}

\subsection{Obtaining the scaling law \label{obtaining_scaling_law}}
\begin{figure}[h!]
\includegraphics[scale=1]{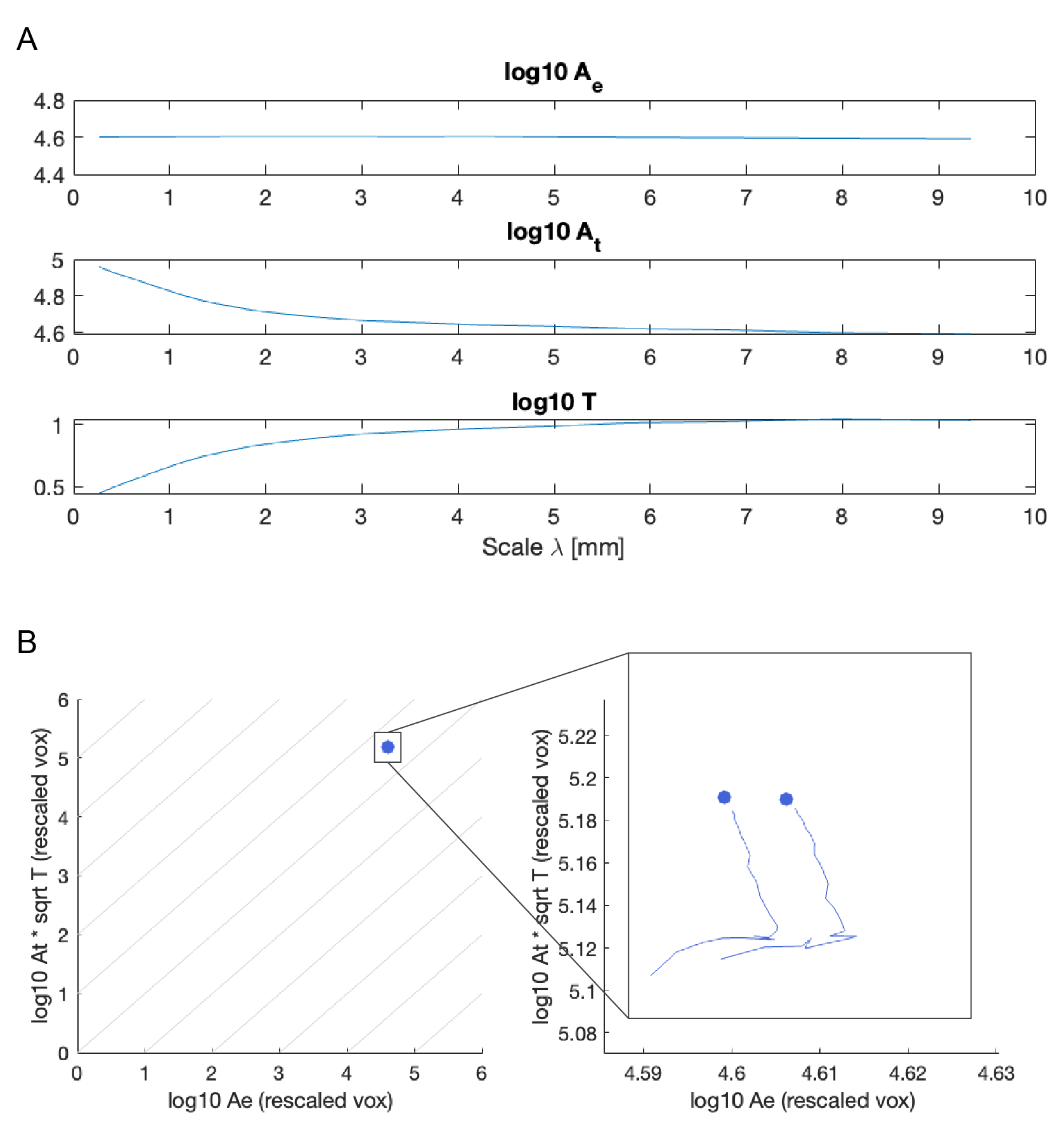}
\caption{\textbf{Unscaled quantities cannot be used to verify scaling law.}\label{Irescaling} \textbf{A:} The exposed surface area $A_e$ barely changes across scales, and $A_t$ only changes minimally. \textbf{B:} Plotted in the scaling law plane (as Fig.~\ref{fig:scaling}) the data points barely have any variance, and overlap each other substantially. Even after zooming in, the trace for the the coarse-graining procedure (solid line) is virtually a vertical line (with a small artifactual `tail' coming from very large voxel sizes).}
\label{sfig:constAt}
\end{figure}

From the coarse-grain procedure (Fig.~\ref{fig:meltingprocess}), we obtain surface meshes for the pial and white matter surface at each spatial scale. From those, we derive exposed area, total pial surface area, and average cortical thickness as described in Methods. However, the coarse-graining procedure, by itself, barely changes the exposed surface area, and only changes the total surface area minimally (Fig.~\ref{sfig:constAt} A). As the voxel size changes at each scale, it only starts to affect the exposed area at very large scales, where effectively the voxels are no longer a good description of the shape of the skull. Thus, if we scatter the raw data points from the coarse-graining procedure in the plane of the scaling law, there is barely any variance in the data (Fig.~\ref{sfig:constAt} B). Even after zooming in, we see an almost vertical line (Fig.~\ref{sfig:constAt} B). This is expected and is not evidence for or against our hypothesis.

Instead, we need to transform the data into a perspective that makes sense to be be seen in the scaling law plane. Instead of thinking of the coarse-graining procedure as a process that re-renders the cortex at increasing voxel sizes, we can instead think of the procedure as rescaling the original cortical mesh into increasingly smaller sizes, and re-rendering it with the same fixed voxel size. The analogy using Britain's coastline is that instead of measuring the coastline with increasingly smaller rulers, we resize the map of the coastline to increasingly smaller sizes, but keep the size of the ruler the same. Both procedures are equivalent and produce the same fractal dimension.

To achieve this procedure, we re-scaled our measurements of $A_e$, $T$ and $A_t$ by the voxel size ($\lambda$) and a fixed factor ($l_r$):
\begin{equation}
\begin{split}
    A_e^r & = \frac{A_e}{(\lambda \times l_r)^2}\\
    A_t^r & = \frac{A_t}{(\lambda \times l_r)^2}\\
    T^r & = \frac{T}{(\lambda \times l_r)}.\\
    \end{split}
\end{equation}

As we used isometric cubes as voxels, the voxel size refers to the length of a single side. E.g. $\lambda=1$ if we used a isometric $1 \times 1 \times 1 mm^3$ voxel.

$l_r$ is a fixed factor for each cortical hemisphere, and does not change with $\lambda$. We use it to systematically shift all the data points within a range, such that the re-scaled quantities are not larger than those from the original cortical meshes. One can easily verify that $l_r$ will not change the slope or offset of any scaling law, but simply represents a constant shift to all data points. In our data, some of the re-scaled quantities would indeed be larger than those from the original cortical meshes, as the voxel size we choose is limited at the smaller end only by computational resources. In other words, we can use very small voxel sizes (relative to the mesh), which after re-sizing would yield very large values of $A_e^r$, $A_t^r$, and $T^r$. To avoid  this, we chose $l_r$ simply as the ratio of the $I$ (isometric term) of the mesh at the smallest scale we used relative to the original mesh, divided by the $\lambda$ of the smallest scale:

\begin{equation}
l_r=\frac{1}{\lambda_s} \times \frac{I_s}{I_o},
\end{equation}
where $\lambda_s$ is the smallest voxel size used for a particular cortex, $I_s$ is the corresponding $I$ for this cortex at the smallest voxel size.

Finally, $I_o$ is the $I$ term for the original cortical mesh. Indeed, the ratio of $\frac{I_s}{I_o}$ is always close to, but larger than one in our dataset. Thus, we can see in Fig.~\ref{fig:scaling} A that most traces start very close to the original data point, indicating that our finest scale is reconstructing the original surfaces well.

Note that the re-scaling is isometric, meaning that is only affects $I$, but does not change the data in the $K \times S$ plane (Fig.~\ref{fig:KxS}). Rescaling by a factor proportional to $\lambda^2$ can therefore be understood as distributing the data points along the $I$ axis, while $l_r$ can be understood as fixing the position in the $I$ axis relative to the $I$ of the original mesh.

Fig.~\ref{fig:scaling} A is produced with these rescaled quantities as described above. The only final step in producing Fig.~\ref{fig:scaling} A, and in calculating the associated slopes, is the removal of artifactual data points where $\frac{A_t}{A_e} < 0$, which can occur at very large voxel sizes relative to the cortical mesh.

The algorithmic implementation in MATLAB can be found on Github:
\url{https://github.com/cnnp-lab/CorticalFoldingAnalysisTools/blob/master/Scales/fastEstimateScale.m}, as part of our Cortical Folding Analysis MATLAB package \url{https://github.com/cnnp-lab/CorticalFoldingAnalysisTools/}, which also has been recently updated with a graphical user interface.

\subsection{Species-specific details\label{species-specific}}
\begin{figure}[h!]
\vspace{-1cm}
\includegraphics[width=0.9\textwidth]{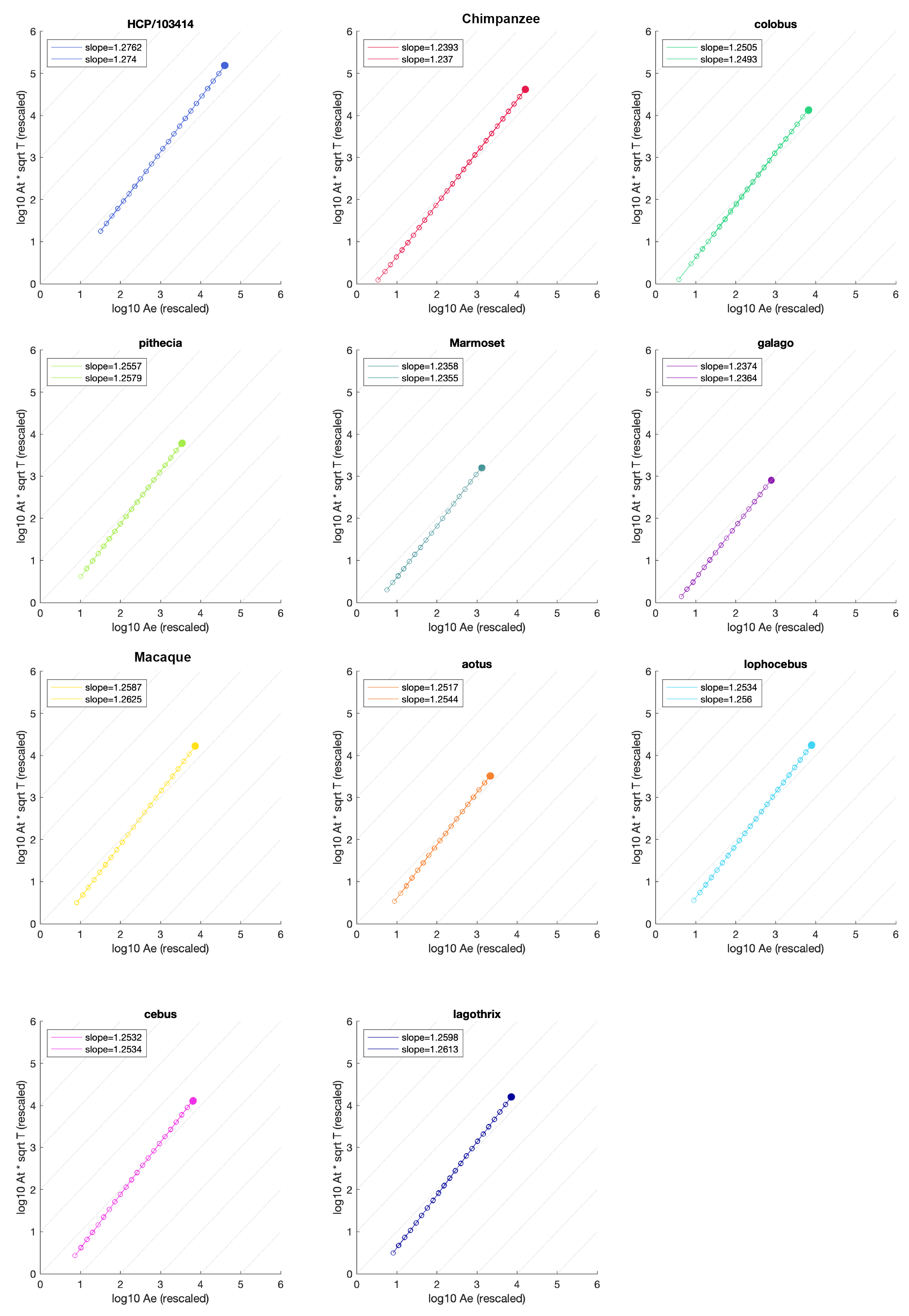}
\caption{\textbf{Detailed scaling plots for each species.} Filled circle is the original mesh, empty circle are data points from the coarse-graining algorithm. Empty circles are connected by a line for visualisation. Two hemispheres are analysed separated for each species, although data points overlap substantially in plot. Slope estimates are given at top left corner in each plot. \label{sfig:allspec}}
\end{figure}

In the following, we will show the detailed data for each species in terms of their scaling behaviour in Fig.~\ref{sfig:allspec}. Videos for all species, showing the pial surface at each scale can be found under \url{https://bit.ly/3CDoqZQ} for review purposes. Final versions of all underlying data, analysis code, and videos will be published on Zenodo and eLife upon acceptance of the paper.

Given the demonstrated overlap between species, and if their cortices are all approximations of the same form, how can one tell apart their cortices? The answer is that all approximations have a range of validity, which varies between cortices. For a gyrified cortex of area $A_t$\footnote{When no dependence on $\lambda$ is indicated then we are referring to the values for the original cortex}, the coarse-graining will remove details of ever-increasing scale, lowering $A_t(\lambda)$ until attaining lissencephaly for $A_e(\lambda_{lys}) = A_t(\lambda_{lys})$. Indeed, it follows from rewriting Eqn. (\ref{eq:scaling}) in the form of Eqn. (\ref{eqn2_A0}) that a given cortex' shape will be comprised of self-similar structures with areas ranging from $A_0 = g^{-5} A_t$  at their smallest to $A_t$ at their largest, where $g$ is the gyrification index $g =\frac{A_t}{A_e}$. Whenever this self-similar scaling is valid, $A_0$ acquires a further interpretation as the typical size of the smallest structures in a cortex: patches smaller than that must be approximately smooth. Consequently, $N_{structures} \simeq \frac{A_t}{A_0} = g^5$ estimates the number of morphological features in each cortex (and adds a new interpretation for the gyrification index). For example, we estimate the human in our dataset to have about 105 morphological features in each cortical hemisphere, and the galago to have about 4 such morphological features. The corresponding $N_{structures}(\lambda)$ estimates how this number changes over coarse-graining. Suppl.~\ref{morpho} later provides a more detailed discussion on this topic.

\section{Morphometric relations and the range of validity of the fractal approximation \label{morpho}}
%The linear algebra of cortical morphology

There are many geometrical quantities that describe aspects of entire cortical hemispheres. But most information contained in such summary morphometric measures is captured, exactly or to good approximations, by the three used in Eqn (\ref{eqn1_scalinglaw}): the pial (or total) surface area $A_t$, exposed surface area $A_e$, and average cortical thickness $T$. Other quantities, like the gyrification index $g = \frac{A_t}{A_e}$, or the Grey Matter and Total volumes, can be expressed as products of power laws of these quantities; and the logarithm of products of power laws are linear combinations of the logarithms of the constituting variables, with their exponents as coefficients. Thus, the morphology of a given cortex can be fairly summarized as point in a 3-dimensional morphometric space with components given by $\log A_t$, $\log A_e$ and $\log T^2$ (the last exponent guarantees all axes have the same dimension of $\log[Area]$).

This logarithmic structure guarantees that the quantities denoted by products of power laws of the original variables correspond to vectors in the morphometric space: the increase and decrease of their values occur along these vectors, and the planes perpendicular to each vector denote all configurations with the same value for the associated variable. For example, $G = \log g = \log A_t - \log A_e$, the point in the morphometric space associated with the gyrification ratio value $\frac{A_t}{A_e}$, defines a displacement along a vector with coefficients $\{1,-1,0\}$.

The quantities $K$, $S$ and $I$ used in Sec. \ref{results2} likewise correspond to displacements along specific directions in morphometric space, which are all orthogonal to one another. These can be normalized, so that the sum of their squared coefficients equals unity. In this case, expressing cortical shape in the morphometric space in terms of the original variables or the new variables amounts simply to a rotation: a change of orthonormal base.

Figs. \ref{fig:scaling} and \ref{fig:KxS} can be then regarded as particular 2D snapshots, taken from a certain direction, of a 3D set of points.

Of particular note, those vectors with a sum of coefficients equal to zero correspond to dimensionless variables. Any such vector is perpendicular to the $I$ direction, and lie on the $K \times S$ plane.

In this framework, we can picture the step-wise coarse-graining of a cortex, as described in Sec. \ref{results1}, as a trajectory in this morphometric space. Empirically, Sec. \ref{results1}, as codified by Eqn. \ref{eq:scaling}, finds that the trajectories in morphometric space for all studied primates are linear, and largely overlap along a line of constant $K$. For cortical morphology, the first fact implies scale-invariance, the second implies universality.

Aging, development, and the progression of neurodegenerative conditions likewise each correspond to different morphometric trajectories.

Now consider a morphometric trajectory, corresponding to the coarse-graining of a cortex up to the point of lissencephaly, i.e., to the spatial scale $\lambda_{lys}$ such that $A_t(\lambda_{lys}) = A_e(\lambda_{lys})$. In the idealized case (but very close to empirical, as seen in Sec. \ref{results1}), all points along the trajectory should follow Eqn (\ref{eq:scaling}), starting at the original values for the morphometric measures and ending at the intersection of the $K=constant$ coarse-graining trajectory and the $g=1$ line that defines the limit for lissencephaly (points with $g<1$ are geometrically possible, but are largely avoided by the trajectories of actual cortices, although not by those of all objects; see \ref{sfig:nonbrainvalid}). This trajectory spans a range of values for each morphometric measure: for dimensionless morphometric measures, such span is fully specified by Eqn (\ref{eq:scaling}). For all others, one needs also to specify the voxel rescaling. The scheme described in Sec.\ref{Irescaling} simply guarantees that, for a constant $k = A_t A_e^{-\frac{5}{4}}T^\frac{1}{2}$, the fundamental area element $A_0 = \frac{T^2}{k^4} = \frac{A_e^5}{A_t^4}$ (and thus also the average thickness $T$) is kept constant for all realisations of the coarse-graining process. We can regard these as approximations of the same original cortex, all drawn with the same resolution, but in ever smaller isometric sizes.

This choice of spatial scale then allows our identification of $A_0$ as the typical size of the smallest morphological features in a given cortex. Equivalently, it is also the area of the largest possible lissencephalic cortex for a given average cortical thickness, just at the cusp of the onset of gyrencephaly.

This same choice also enable us to easily compute the expected span of the various morphometric measures over coarse-graining, assuming the universal scaling: each will range between its original value (expressed as a function of the original values for $A_t$, $A_e$ and $T$), and its value at spatial scale $\lambda_{lys}$, where  $A_t(\lambda_{lys}) = A_e(\lambda_{lys}) = A_0$: we simply replace $A_t$ and $A_e$ by $A_0$, and $T$ by $\sqrt{A_0 k^4}$. The ratios between the initial and final values of all morphometric measures will be given as powers of $g$, imbuing the gyrification index with a new significance.

We list below the exspans for all such morphometric measures ($S = \log s$, $K = \log k$ and $I = \log v_I$) 

\begin{table}[!h]
\begin{tabular}{lllll}
Morphometric measure    & Original Value =                                                          & Span      $\times$    & Value at lissencephalic limit &  \\
$T^2$           & $T^2$                                                                     & $1$                   & $A_0 k^4$                     &  \\
$A_0$           & $\frac{Ae^5}{At^4}$                                                       & $1$                   & $A_0$                         &  \\
$A_e$           & $A_e$                                                                     & $g^4$                 & $A_0$                         &  \\
$A_t$           & $A_t$                                                                     & $g^5$                 & $A_0$                         &  \\
$V_{total}$     & $\approx\frac{2}{9\sqrt{3\pi}}A_e^{\frac{3}{2}}$                          & $g^6$                 & $\approx\frac{2}{9\sqrt{3\pi}} A_0^{\frac{3}{2}}$ &  \\
$V_{GM}$        & $A_t T$                                                                   & $g^5$                 & $A_0^{\frac{3}{2}}k^2$        &  \\
$g$             & $\frac{At}{Ae}$                                                           & $g$                   & $1$                           &  \\
$N_{features}$  & $\approx\frac{At^5}{Ae^5}$                                                & $g^5$                 & $1$                           &  \\
$k  $           & $A_t A_e^{-\frac{5}{4}}T^\frac{1}{2}$                                     & $1$                   & $k$                           &  \\
$s  $           & $A_t^{\frac{3}{2}}A_e^{\frac{3}{4}}T^{-\frac{9}{2}}$                      & $g^\frac{21}{2}$      & $k^{-9}$                      &  \\
$v_I$           & $A_t A_e T^2$                                                             & $g^9$                 & $A_0^3 k^4$                   &  \\
$\hat{k}$       & $A_t^{\frac{4}{\sqrt{42}}}A_e^{-\frac{5}{\sqrt{42}}}T^\frac{2}{\sqrt{42}}$& $1$                   & $\hat{k}$                     &  \\
$\hat{s}  $     & $A_t^{\frac{2}{\sqrt{14}}}A_e^{\frac{1}{\sqrt{14}}}T^{-\frac{6}{\sqrt{14}}}$& $g^\frac{18}{\sqrt{14}}$ & $\hat{k}^{-3\sqrt{3}}$    &  \\
$\hat{v}_I$     & $A_t^{\frac{1}{\sqrt{3}}}A_e^{\frac{1}{\sqrt{3}}}T^\frac{2}{\sqrt{3}}$    & $g^{3\sqrt{3}}$       & $A_0 \hat{k}^{\sqrt{14}}$     &  \\
                &                                                                           &                       &                               & 
%$s  $          & $A_tA_e^2T^{-3}$                                                          & $g^7$                 & $k^{-6}$                      &  \\  %Alternative definition of s (s_old^(2/3)
\end{tabular}
\end{table}

\section{Validation with non-brain objects \label{validation_nonbrain}}
We chose to apply our coarse-graining procedure to a range of objects to validate our algorithm, and we included both negative and positive controls. 

\begin{figure}[h!]
\includegraphics[width=\textwidth]{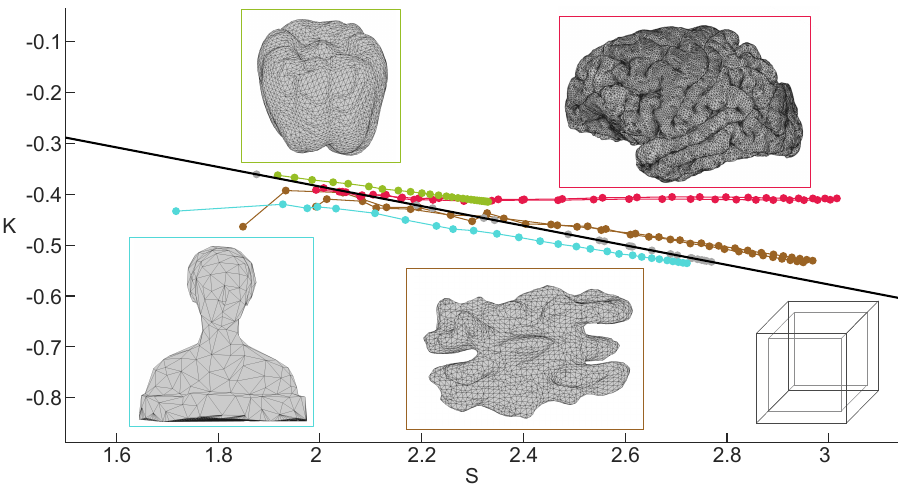}
\caption{\textbf{Scaling behaviour of various objects in $K \times S$ space.} Thick black line indicates $K=-\frac{1}{9}S$ and a gyrification index $g=\frac{A_t}{A_e}=1$. Points above and below the line have, respectively, $g>1$ and $g<1$. Colour-coded boxes show the surface of the objects we analysed. For simplicity we show the outer ``pial'' surface of each object. In the case of the box, we indicated both outer and inner surfaces. Green data points correspond to the bell pepper, red to the human brain, cyan to the `Laurana' bust, brown to the walnut, and grey to the box of finite thickness.  \label{sfig:nonbrainvalid}}
\end{figure}

As a first positive control, we used a simple box with finite thickness (i.e. we simulated a inner ``white matter'' surface and an outer ``grey matter'' surface, both being a cube, one positioned inside the other). A schematic is shown in Fig.~\ref{sfig:nonbrainvalid}. We know based on theoretical consideration that this box must be aligning exactly with the $K=-\frac{1}{9}S$ line, as its $A_t$ must be the same as its $A_e$ at every stage of coarse-graining, which also implies a ``fractal dimension'' of 1. This is exactly the case in our plot of $K$ against $S$, as all the grey data points from the coarse-graining algorithm align on $K=-\frac{1}{9}S$ (thick black line).

As negative controls, we included three non-brain objects, a bell pepper, two walnut halves, and a coarse outline of a bust/figurine known as Laurana. All three of these objects show a self-similar, or fractal, regime (corresponding to partially straight trajectories in $K \times S$) in our coarse-graining procedure, but their trajectories in $K \times S$ space are not flat (Fig.~\ref{sfig:nonbrainvalid}), meaning their fractal dimensions are distinct from 2.5. Their trajectories also do not overlap with each other, indicating that they are fundamentally different shapes.

The human brain is included here as a reference and shows a clear flat trajectory (Fig.~\ref{sfig:nonbrainvalid}) as presented in the main text. Also worth noting, after the human trajectory intersects the $K=-\frac{1}{9}S$ line, it veers off to follow the line closely, indicating that the $A_e$ and $A_t$ remain the same with further steps of coarse-graining. Effectively, the human brain transitions to be lissencephalic convex structures once their fractal regime ends. For simplicity, we excluded these lissencephalic data points of extreme coarse-graining from the results in the main text.

\section{Validation with randomising grid for coarse-graining \label{validation_jitter}}
\begin{figure}[h!]
\includegraphics[width=\textwidth]{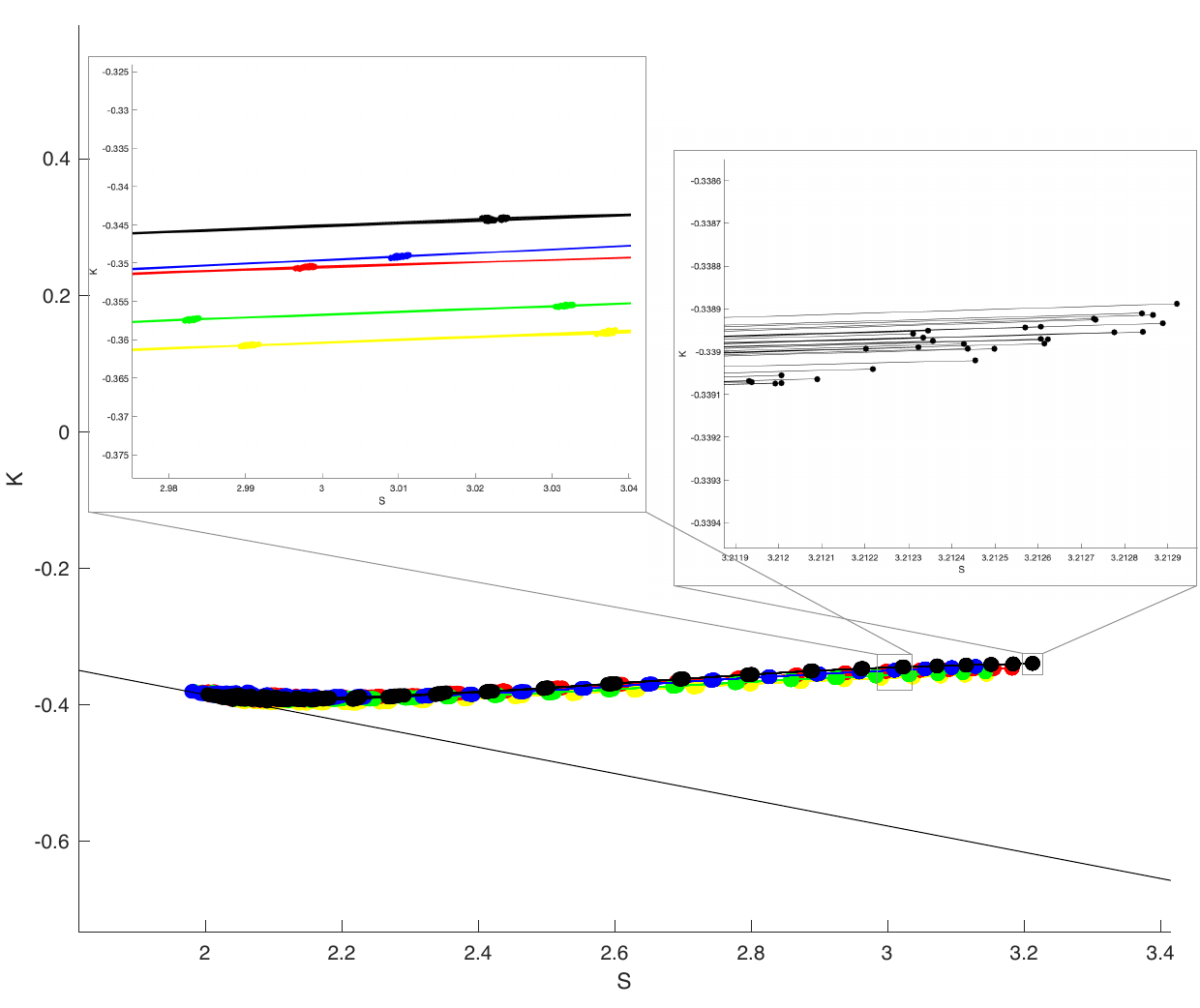}
\caption{\textbf{Randomising grid for coarse-graining yields very consistent results within individuals in $K \times S$ space.} Thick black line indicates $K=-\frac{1}{9}S$ and a gyrification index $g=\frac{A_t}{A_e}=1$. Each colour in the plot indicates a different human individual (HCP 103414 - yellow, 135225 - red, 138534 - green, 144832 - blue , 148840 - black, respectively). 30 jittered grid outputs are shown for each individual, lines connect datapoints of the same jittered version. Zoomed in panels show that the jittered outputs have far less variation within than between individuals.\label{sfig:jitter}}
\end{figure}

To assess robustness of the coarse-graining algorithm, we ran 30 different realisations of the algorithm, but with a small random shift of the grid position relative to the surface meshes. The shift is chosen to be within the radius of $\lambda$. This allows subtle changes in the voxelisation at the boundary of grey matter and white matter. Interestingly, over five different human individuals, we observed subtle changes in the values $S$ over the 30 realisations, and to a lesser degree $K$ (Fig.~\ref{sfig:jitter}). Additionally, the variation between individuals, especially in $K$ at smaller spatial scales, is far greater than the variation introduced by the jittered realisations. These results suggest that the coarse-graining algorithm is extremely robust towards subtle changes at the grey or white matter boundary, especially far away from the lissencepahlic limit.

\section{Ageing process \label{ageing}}
%In the main text, we used the ageing process as an example for scale-dependent biological process. In our particular example, the dramatic scale-dependency of the effect of $K$ can be visually and intuitively understood by looking at changes in total surface area (Fig.~\ref{fig:ageing}~B, also see Suppl.~\ref{ageing_allvar}): Fig.~\ref{fig:ageing}~C shows some coronal slices of the cortical surface. At scale 0.27~mm, the gyri in the younger subjects are densely packed, but the older subjects show the expected (see e.g.  \citep{Jin2018,Madan2019} for recent investigations and references therein) widening between gyral walls and decrease in gyral surface area at the crown. At 1.86mm, the younger cortices have already partially ``melted'', erasing most small sulci between the densely packed gyri. In the older humans, however, the gyri are less dense, the sulci more open, and thus, at 1.86mm, most gyri and sulci have not been erased yet.  At scale 7.94mm, both young and old brains have ``melted'' down to very similar near-lissencephalic shapes. 

%More broadly, one can regard the melting process as a way of determining how cortical area is allocated across different scales. As the cortex `melts', the contributions to the total area from features smaller than the cut-off scale are eliminated. In the example in Fig.~\ref{fig:ageing}, for instance, we can say about half ($10^{5-4.7}=10^{0.3} \approx 2$) the total area in the 80 y.o. cortices is present in features smaller than 4mm. 

\subsection{All morphometric variables\label{ageing_allvar}}

\begin{figure}[h!]
\includegraphics[width=0.8\textwidth]{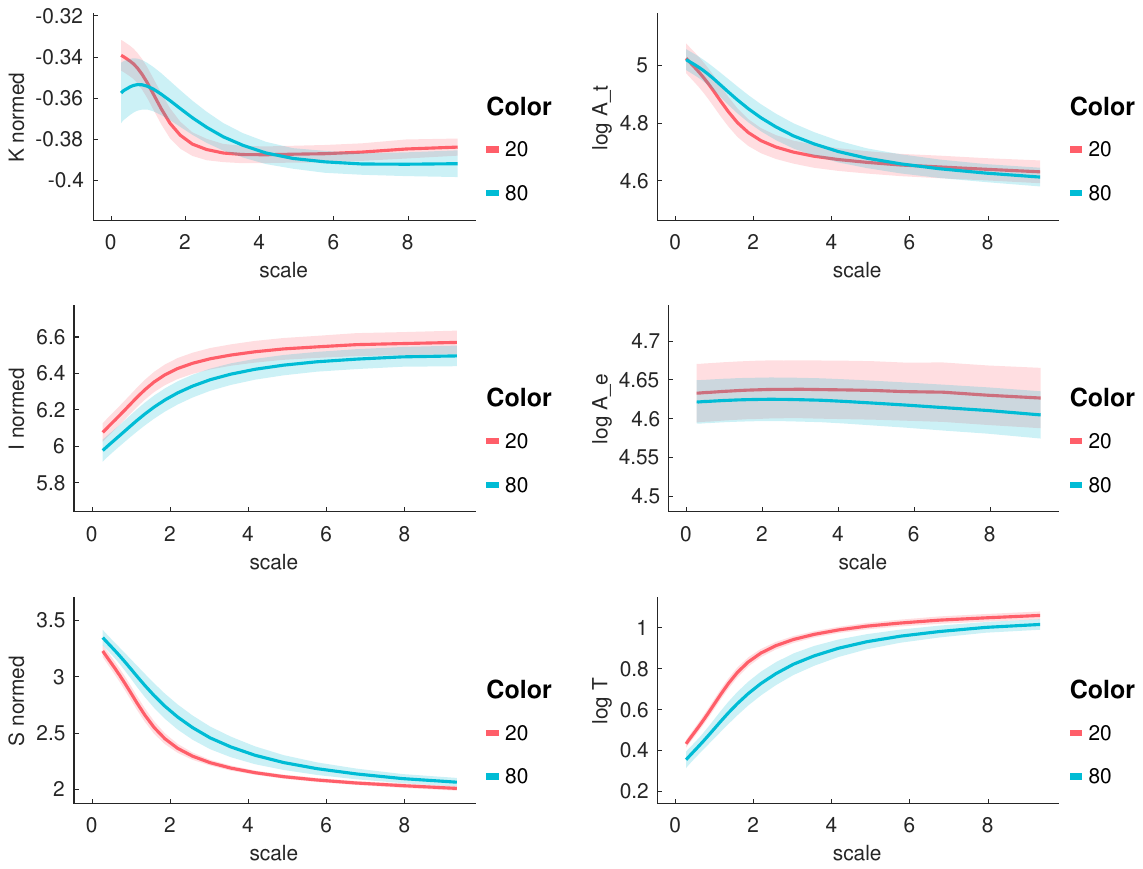}
\caption{\textbf{All morphological variables across scales in the 20 y.o. and 80 y.o. cohort.} Each panel shows a morphological metric, and data is shown for a group of 20-year-olds (red, n=27) and a group of 80-year-olds (blue, n=86). Mean and standard deviation are shown as the solid line and the shaded area respectively.} \label{sfig:ageing_data}
\end{figure}

\begin{figure}[h!]
\includegraphics[width=0.8\textwidth]{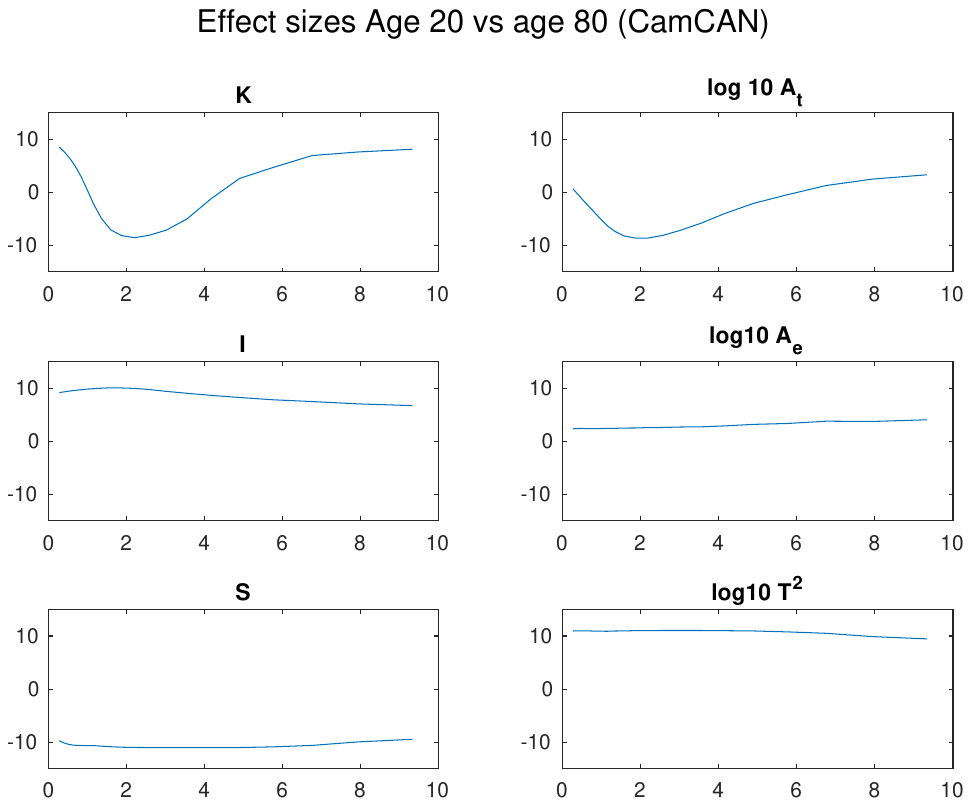}
\caption{\textbf{Effect size between the 20 y.o. and 80 y.o. cohort in all morphological variables.} Each panel shows a morphological metric, and data is shown for a group of 20-year-olds (red, n=27) and a group of 80-year-olds (blue, n=86). Effect size is measured as the ranksum z statistic between the two groups.} \label{sfig:ageing_effect}
\end{figure}

In the main text, we used the ageing process as an example for scale-dependent biological process.
For completeness, here we also show all the morphometric variables across scales in the 20 y.o. and 80 y.o. cohort in Fig. \ref{sfig:ageing_data}, and the corresponding effect sizes in Fig. \ref{sfig:ageing_effect}. As expected, $A_e$ shows very little difference between the two cohorts. $I$, $S$, and $T$ show some effects at scales smaller than 4~mm, but the effect decreases monotonously for higher scales. $K$ demonstrates a more complex scale-dependent effect: larger $K$ in younger subjects at small scales of 0.25~mm (in agreement with previous native scale analyses \cite{pnas2016}), smaller $K$ in younger subjects at  scales of approx. 2~mm, and a return to larger $K$ in younger subjects at large scales of more than approx. 5~mm. Note, however, that despite these large effect sizes, the actual change of values of $K$ is within the range of variation expected across species (seen in the main text). The range of variation of $K$ is approx 0.04 here, and at least an order of magnitude smaller than the range of variation in $S$ (approx 1.5).

\subsection{Confirmation in independent dataset \label{ageing_nki}}
\begin{figure}[h!]
\includegraphics[width=\textwidth]{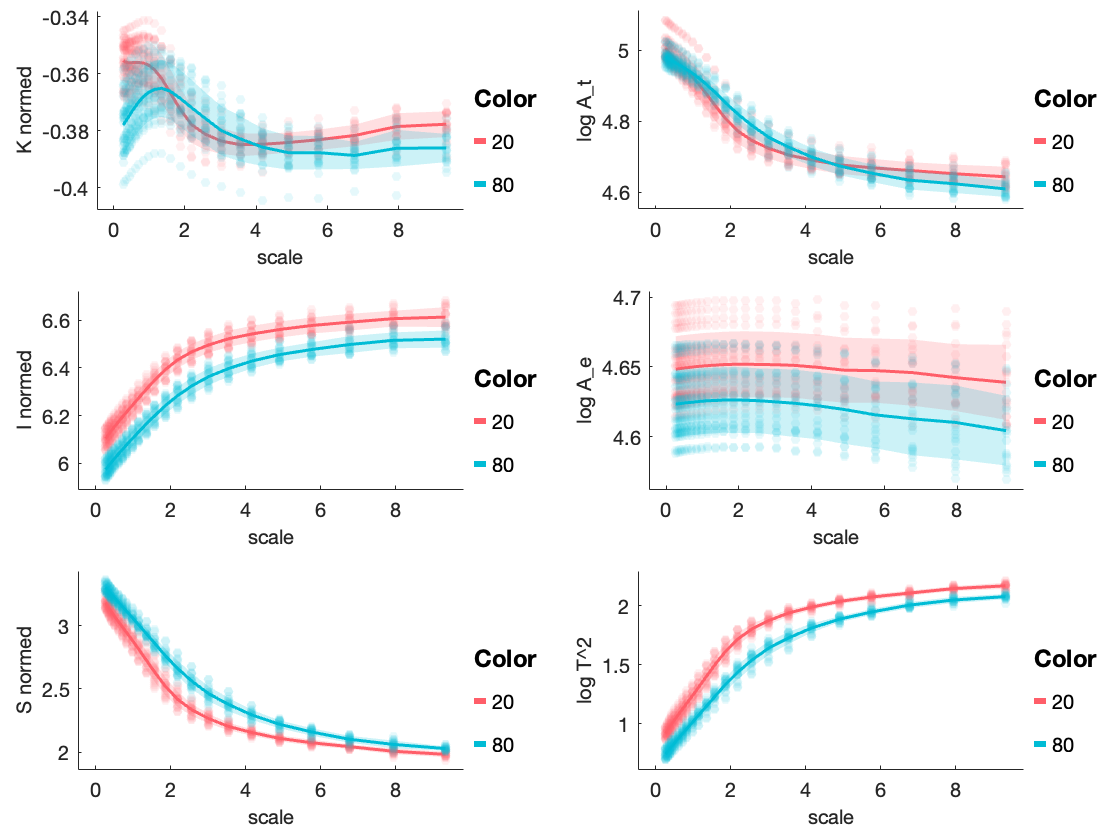}
\caption{\textbf{All morphological variables across scales in the 20 y.o. and 80 y.o. cohort in a separate (NKI) dataset.} Each panel shows a morphological metric, and data from an independent dataset (NKI) is shown for a group of 20-year-olds (red, n=10) and a group of 80-year-olds (blue, n=10).} \label{sfig:ageing_data_NKI}
\end{figure}

\begin{figure}[h!]
\includegraphics[width=\textwidth]{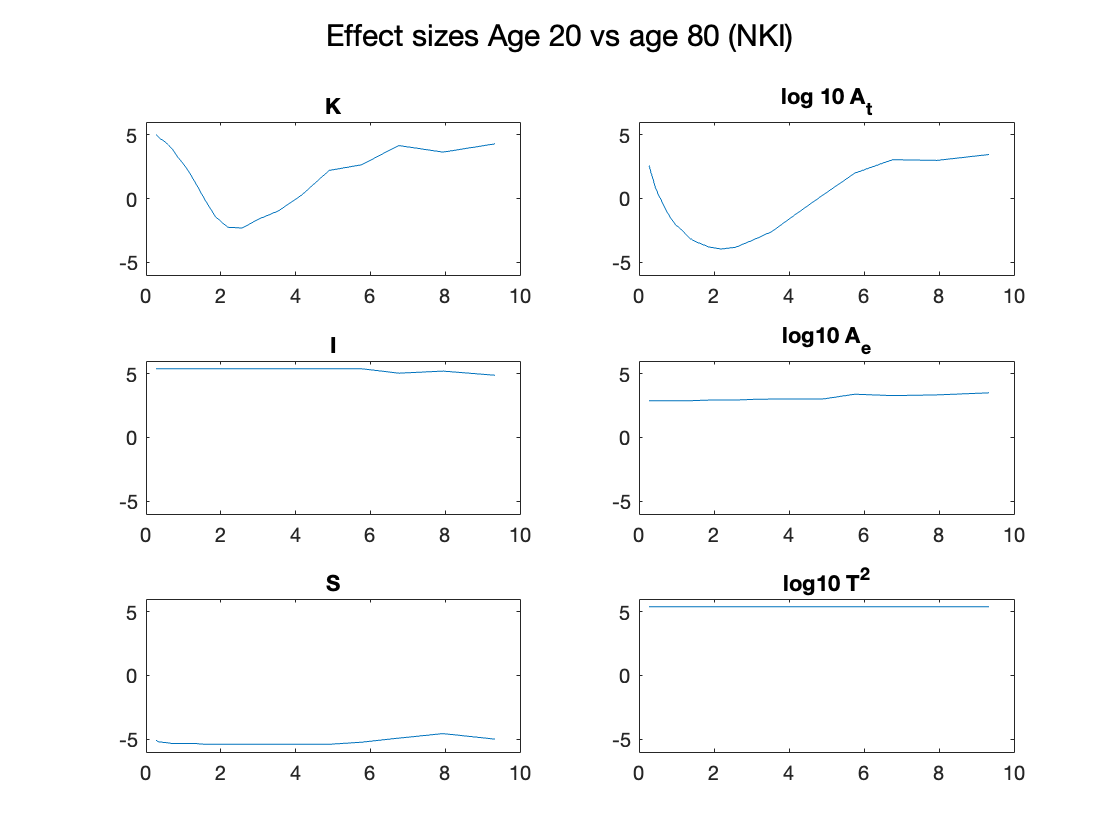}
\caption{\textbf{Effect size between the 20 y.o. and 80 y.o. cohort in all morphological variables in a separate (NKI) dataset.} Effect size is measured as the ranksum z statistic between the two groups. Mean and standard deviation are shown as the solid line and the shaded area respectively.} \label{sfig:ageing_effect_NKI}
\end{figure}

To confirm that our observed ageing effect was not driven by sample or data-specific properties, we also analysed an independent dataset (NKI) with the same methods. Here, we show the equivalent figures to Fig.~\ref{sfig:ageing_data} and Fig.~\ref{sfig:ageing_effect} for the NKI data in Fig.~\ref{sfig:ageing_data_NKI} and Fig.~\ref{sfig:ageing_effect_NKI}. The same qualitative patterns can be seen in all plots, including the scale-specific effects in $K$ and $A_e$ at around 2~mm.

\end{document}